%% file: article.tex
\documentclass[final]{siamart1116}

\input{ex_shared}

\ifpdf
\hypersetup{
  pdftitle={\TheTitle},
  pdfauthor={\TheAuthors}
}
\fi

\begin{document}

\maketitle


\begin{abstract}
In this work, we extend and analyze the nonperturbative  Maxwell-Schr\"odinger-Plasma (MASP) model. This model was proposed to describe the high order optical nonlinearities, and the low density free electron plasma generated by a laser pulse propagating in a gas. The MASP model is based on nonasymptotic, ab-initio equations, and accurately uses self-consistent description of micro (quantum)- and macro (field)- variables. However, its major drawback is a high computational cost, which in practice means that only short propagation lengths can be calculated. In order to reduce this cost, we study the MASP models enriched by a macroscopic evolution equation for polarization, from its simplest version in a form of transport equation, to more complex nonlinear variants. We show that homogeneous transport equation is a more universal tool to simulate the high harmonic spectra at shorter times and/or at a lower computational cost, while the nonlinear equation could be useful for modeling the pulse profiles when the ionization level is moderate. The gain associated with the considered modifications of the MASP model, being expressed in reduction of computational time and the number of processors involved, is of 2-3 orders of magnitude.
\end{abstract}



\input intro
\input statement

\input EandU
\input parallel
\input results

\input conclusion

\section*{Acknowledgements} The authors thank Compute Canada for access to high performance computer facilities for our simulations in highly nonlinear optics.

\bibliographystyle{siamplain}
\bibliography{refs}

\end{document}

%% file: ex_shared.tex
\usepackage{braket,amsfonts,amsopn}
\usepackage{graphicx}
\usepackage{epstopdf}
\usepackage{algorithmic}
\usepackage{array}
\usepackage{comment}
\usepackage[caption=false]{subfig}

\ifpdf
  \DeclareGraphicsExtensions{.eps,.pdf,.png,.jpg}
\else
  \DeclareGraphicsExtensions{.eps}
\fi

\numberwithin{theorem}{section}

\newsiamremark{rem}{Remark}

\newcommand{\TheTitle}{Nonlinear Nonperturbative Optics Model Enriched by Evolution Equation for Polarization}
\newcommand{\TheAuthors}{M. Lytova, E. Lorin, and A. D. Bandrauk}

\headers{Nonlinear Nonperturbative Optics Model Enriched...}{\TheAuthors}

\title{{\TheTitle}\thanks{Submitted to the editors \today.
}}

\author{
  Marianna Lytova%
  \thanks{School of Mathematics and Statistics, Carleton University, Ottawa, Canada, K1S 5B6
    \mbox{(\email{marianna.lytova@carleton.ca}, \email{elorin@math.carleton.ca})}.}%
  \and
  Emmanuel Lorin%
  \footnotemark[2]
  \thanks{Centre de Recherches Math\'{e}matiques, Universit\'{e} de Montr\'{e}al, Montr\'{e}al, Canada, H3T~1J4.}
  \and
  {Andr\'e D. Bandrauk
  \thanks{L\MakeLowercase{aboratoire de chimie th\'{e}orique,} F\MakeLowercase{acult\'{e} des} S\MakeLowercase{ciences,} U\MakeLowercase{niversit\'{e} de} S\MakeLowercase{herbrooke,} S\MakeLowercase{herbrooke,} C\MakeLowercase{anada,} J1K 2R1 \MakeLowercase{(\email{andre.bandrauk@usherbrooke.ca}).}}} \footnotemark[3]
}


%% file: intro.tex
\section{Introduction}
The discovery of attosecond pulses \cite{Ferray,McPherson} and of laser filaments \cite{Mourou}  has led to the need for an appropriate mathematical description of the evolution of ultrashort, that is only few-optical-cycles, laser pulses. Perturbative methods based on the approximate equations, e.g. Nonlinear Schr\"odinger Equation (NLSE)-type, although advantageous for not too short pulses, have essential intrinsic restrictions when applied to ultrashort pulses. These issues were addressed in some key papers \cite{13b,Husakou,kalosha}. Over the years, several advanced models were proposed, and we refer to a set of review papers \cite{99,12,Kolesik2014}. Certain complex models, e.g. including High Order Kerr Effect (HOKE) and Unidirectional Pulse Propagation Equation (UPPE) \cite{Babushkin2014,Kolesik2}), allow for accurate simulations and analysis of laser filamentation in some physical regimes \cite{Babushkin2014,21,Champeaux2008,kol2,Skupin200614,berg2}. However, an appropriate modeling of free-electron plasma generated by the ultrashort intense laser pulse, as well as of nonperturbative evolution of the pulse in nonlinear media, is still an open problem that is only partially addressed \cite{B2013,bejot,K2013,new_last,mol2, Richter2013,Spott2014,strelkov}.\\
\indent Now we briefly specify the principal features of the standard models used to study filament propagation \cite{BLM,99,Kolesik2014}. In a broad sense, there are two general standard approaches to describe short-pulse propagation in a nonlinear medium \cite{newell2,kolesik2016}:
\begin{itemize}
  \item models considering the evolution of the envelope of the electromagnetic wave. As
      a rule, they are based on time-dependent NLSE, and the second
      generation equations, e.g. the Nonlinear Envelope Equation (NEE)
      \cite{brabec};
  \item models based on full electric field propagators. It is the set of
      equations connected with Korteveg-de Vries (KdV) equation, e.g. the UPPE \cite{kolesik2016}.
\end{itemize}
Applied to nonlinear optics, both ``parental'' equations, NLSE and KdV, as
well as their descendants, are reductions obtained under some assumptions
from Maxwell's equations. To describe the linear and nonlinear response of the medium, the
perturbative expansion for polarization vector with the susceptibility coefficients
$\chi^{(2m+1)}$ was used \cite{Boyd}. In the strict sense, one can classify standard models as perturbative, even if they account HOKE \cite{99}.\\
\indent Summing up, despite the quite reliable description of the evolution of the ultrafast laser pulses and relatively low computational cost, both standard approaches (i) use approximate but not exact propagation equations; (ii) employ perturbative expansion for the
polarization; (iii) receive the frequency-property of the medium as an external parameters; (iv) compute free electron density phenomenologically \cite{99,Kolesik2014}. All these properties suggest that the solution of the problem can be improved and the involvement of non-standard nonperturbative models will be highly useful.
The fully self-consistent mathematical model must include side by side \cite{2,phyd}:
\begin{itemize}
  \item Maxwell's equations (MEs) modeling the evolution of the laser 
      field under the response of the molecules  in the gas region where
      the pulse propagates;
  \item the set of time-dependent Schr\"odinger equations (TDSEs), which
      describes the interaction of the molecules with an electromagnetic
      field of the pulse;
  \item kinetic equations, in order to take into account the dynamics
      of the generated free-electron plasma.
\end{itemize}
In practice, the realization of such Maxwell-Schr\"odinger-Plasma (MASP) model \cite{2} faces extremely high computational cost for solving the system of TDSEs, and requires high performance computing. Nevertheless, over the past 10 years important results have been achieved within this model, and the studies are in progress, see e.g. \cite{2, cicp,phyd, Lorin2015, Lyt}. Namely, it was suggested  \cite{springer,Lorin2015} to enrich the MASP model by a {\it macroscopic} transport equation modeling the nonperturbative polarization. It was demonstrated that this approach allows to reduce drastically the number of \emph{microscopic} TDSEs in the model thus making the numerical computation of the model much faster. The goals of this paper is (i) to justify mathematically well-posedness of the thus supplemented MASP model, and (ii) to derive more accurate and sophisticated evolution equations. \\
\indent This paper is organized as follows. In Section \ref{sec:model}, we derive the MASP equations enriched by the evolution equations modeling the propagation of the polarization. Section \ref{sec:analysis} is devoted to the mathematical analysis of the enriched model. Numerical approximation and parallel computing aspects are addressed in Section \ref{sec:numerics}. Several numerical experiments are presented in Section \ref{sec:experiments}. We finally conclude in Section \ref{sec:conclusion}. 

%% file: statement.tex
\section{MASP Model and its Development}\label{sec:model}
\begin{figure}[h]
\begin{center}
\center{\includegraphics[scale=0.5,trim=100 100 220 80, clip]{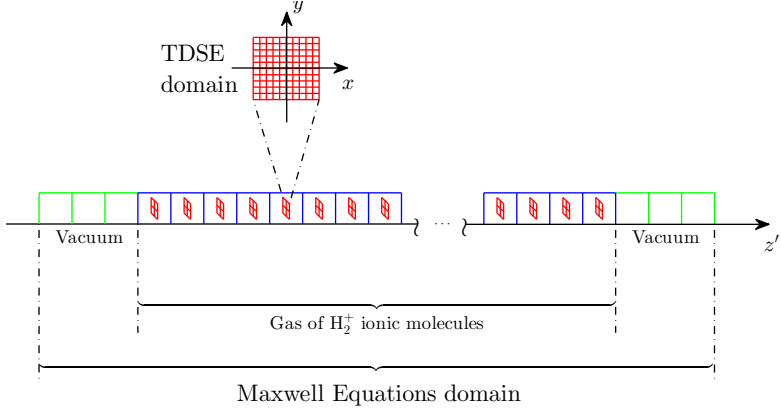}}
\caption{Spatial domain decomposition for the 1d-2d (macro-micro) spatially approximated MASP model.}
\end{center}
\label{Spa_Dom}
\end{figure}

\begin{figure}[b!]
\begin{center}
\includegraphics[scale=0.5,trim=100 140 220 120, clip]{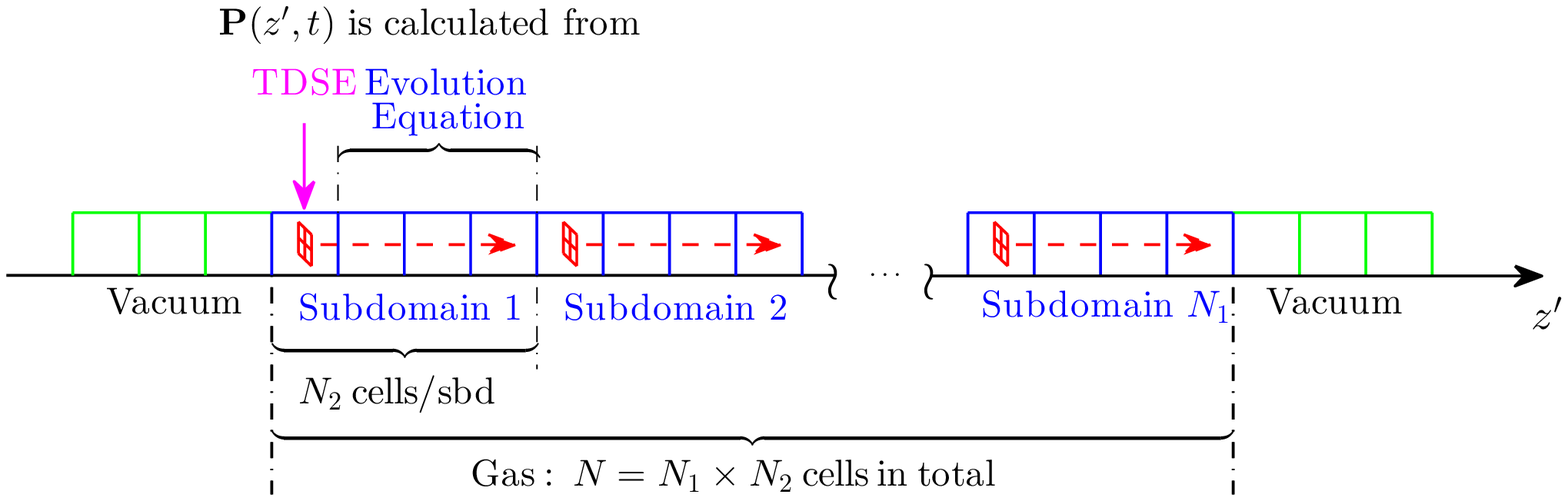}
\caption{Spatial domain decomposition for the 1d-2d MASP model enriched by the evolution equation for polarization.}
\end{center}
\label{fig0}
\end{figure}

The usual spatial setting is as follows: the electromagnetic pulse propagates in sequence through 3 regions: vacuum, molecule-gas and again vacuum. As soon as the wave envelope gets
into the last vacuum region, the electric field components in space and time
are stored. The starting point is a \emph{micro-macro} model constituted by MEs coupled with several TDSEs, describing the nonlinear response of gas subject to an electromagnetic field \cite{2,cicp,phyd}. More precisely, to simulate the \emph{macroscopic} propagation effects, we use the differential form of Ampere-Maxwell's and Faraday-Maxwell's laws in a bounded
spatial domain $\Omega$, with a smooth boundary $\partial\Omega$. We define ${\bf x}'=(x',y',z')^T$ the electromagnetic field space variable, then for nonmagnetic medium, ${\bf B}={\bf H}$:
\begin{subequations}
\begin{align}
\partial_{t} {\bf E}({\bf x}',t) & =  c{\bf \nabla} \times {\bf B}({\bf x}',t) - 4\pi (\partial_t {\bf P}({\bf x}',t) + {\bf J}({\bf x}',t)), \label{AmpLaws}\\
\partial_{t} {\bf B}({\bf x}',t) & =  -c{\bf \nabla} \times {\bf E}({\bf x}',t) \label{FarLaws}.
\end{align}
\end{subequations}
Thus, to compute from \cref{AmpLaws,FarLaws} the electromagnetic field vectors ${\bf E}({\bf x}',t)$ and
${\bf B}({\bf x}',t)$, we need to determine the polarization ${\bf P}({\bf x}',t)$ and the current density ${\bf J}({\bf x}',t)$ vectors from the microscopic equations.\\
\indent We will consider interaction between a molecule, typically $\mathrm{H}_{2}^{+}$ for simplicity,  and the field in the dipole approximation, which is valid when the smallest internal wavelengths $\lambda_{\mathrm{min}}$ of the electromagnetic
field are much larger than the molecule size $\ell$ (in the same direction),
that is $\ell \ll \lambda_{\mathrm{min}}$. Typically we have
$\lambda_{\mathrm{min}} \approx 800$ nm (Ti:Sapphire laser) and for
$\mathrm{H}_{2}^{+}$-molecule size $\ell\approx$ 0.1 nm. The simple molecule $\mathrm{H}_{2}^{+}$ possesses two advantages: it has only one electron, and is symmetric with respect to the point bisecting the
interval between the nucleus. It follows the symmetry of the Hamiltonian with
respect to the electron coordinate inversion: ${\bf x}\longrightarrow-{\bf x}$, which allows to reduce formally the problem of describing the dynamics of 3 particles, to a 2-body problem, and hence the exact Schr\"odinger equation can be numerically considered \cite{cicp}. However, in practice, even more complex molecules like $\mathrm{O}_2$, $\mathrm{N}_2$ could as well be considered within the MASP, using the single active electron approximation (SAE), \cite{AB6}.\\
\indent Although in its whole generality the MASP model for $\mathrm{H}_{2}^{+}$
includes the motion of the 3 particles \cite{cicp}, we will use the Born-Oppenheimer
approximation throughout the article. It is a typical approximation in quantum chemistry, since the time-scale for dynamics of the interatomic electrons  (attoseconds) is much shorter then the time-scale for the nuclei motion (femtoseconds) \cite{chelk98,NewMol,yuan3}.
At the molecular (\emph{microscopic}) scale, we will denote by ${\bf x}=(x,y,z)^T$ the TDSEs space variable (that is for electron wavefunctions). The components of the polarization vector ${\bf P}({\bf x}',t)$, be to substituted into MEs \cref{AmpLaws,FarLaws}, are computed using a trace operator \cite{2}:
\begin{subequations}
\begin{align}
&{\bf P}({\bf x}',t) = \mathcal{N}({\bf x}')\sum_{i=1}^m {\bf P}_i({\bf x}',t)
 =-\mathcal{N}({\bf x}')\sum_{i=1}^m\chi_{\Omega_i}({\bf x}')\displaystyle\int_{\mathbb{R}^3}\psi^*_i({\bf x},t){\bf x}\psi_i({\bf x},t)d{\bf x}, \label{P_schro}\\
&{\tt i}\partial_{t}\psi_i({\bf x},t)  =  -\dfrac{\Delta_{\bf x}}{2}\psi_i({\bf x},t)+ V_C({\bf x})\psi_i({\bf x},t) + {\bf x}\cdot{\bf E}_{{\bf x}'_i}(t)\psi_i({\bf x},t), \:
\forall i \in \{1,..,m\}, \label{schro}
\end{align}
\end{subequations}
where $V_C$ denotes the \emph{static} interaction potential. The density of ionic molecule $\mathrm{H}_{2}^{+}$ is taken to
be smooth in space, and is denoted by $\mathcal{N}({\bf x}')$. It must be
approximately equal to the initial electrons number density
$\mathcal{N}_{e0}({\bf x}')$ \cite{Chen}, and can be assumed constant in time in case of low level of ionization.
For example, let the ionic $\mathrm{H}_{2}^{+}$-molecule be oriented in the plane $(x,y)$ of the selected Cartesian system, then the nuclear potential is written as:
\begin{equation}\label{V_Coulomb_3D}
  \begin{array}{l}
  V_C({\bf x})=-\Big[\Big(x-\dfrac{R_0}{2}\cos\theta\Big)^2+\Big(y-\dfrac{R_0}{2}\sin\theta\Big)^2+z^2 \Big]^{-1/2}-\\
  \hspace*{2cm} -\Big[\Big(x+\dfrac{R_0}{2}\cos\theta\Big)^2+\Big(y+\dfrac{R_0}{2}\sin\theta\Big)^2+z^2 \Big]^{-1/2},
  \end{array}
\end{equation}
where $\theta\in[0^\circ,90^\circ]$ defines the angle between the molecular axis of $\mathrm{H}_{2}^{+}$ and the $x$-axis, with an internuclear distance $R_0$; in our computations it usually equals to 2 atomic units (a.u.), corresponding to $\approx 0.1$ nm, see e.g. \cite{yuan3}.\\
\indent Solving of the TDSEs \cref{schro} provides a complete set of the wavefunctions,
which in its turn allows to evaluate the ionization level of a gas and obtain a
continuum spectrum of free electrons propagating in a laser pulse. We define the spatial domain
$\Omega=\cup_{i=1}^m\Omega_i$, where $\Omega_i$ denotes the
\emph{macroscopic} spatial domain containing a molecule of reference
associated to a wavefunction $\psi_i$, while in \cref{P_schro} ${\bf
P}_i(t)=\chi_{\Omega_i}{\bf d}_i(t)$ denotes the \emph{macroscopic}
polarization in this domain, and the index $i$ is associated to the specific
coordinate of MEs, ${\bf x}'$. Functions $\chi_{\Omega_i}$ are defined as
$\chi\otimes{\bf 1}_{\Omega_i}$ where $\chi$ is a plateau, and ${\bf
1}_{\Omega_i}$ is the characteristic function of $\Omega_i$. Also ${\bf
d}_i(t)$ is a \emph{microscopic} time-dependent dipole moment of a molecule
belonging to $\Omega_i$:
\begin{equation}\label{dipole}
  {\bf d}_i(t)=-\int_{\mathbb{R}^3}\psi_i^*({\bf x},t){\bf x}\psi_i({\bf x},t)d^3{\bf x}.
\end{equation}
\indent In other words, the domain $\Omega_i$ contains $\mathcal{N}({\bf x}')\hbox{vol}(\Omega_i)$ molecules represented by one single wavefunction
$\psi_i$ (under the assumption of a unique pure state \cite{2}). We now
assume that the spatial support of $\psi_i$ is included in a domain $\omega_i
\subset \mathbb{R}^3$. We allow free electrons to reach the boundary $\partial\omega_i$, where we impose
artificial complex potential, see e.g. \cite{ge}.
Thus, in the discussed model, the part of the wavefunction absorbed at
the boundary generates the plasma of free electrons. Finally in \cref{schro}, ${\bf E}_{{\bf
x}'_i}$ denotes the electric field (supposed to be constant within $\omega_i$) in $\Omega_i$.\\
\indent To close the set of equations, the current density evolution equation
can be integrated in the system via the Drude model \cite{99,cicp} on the MEs domain:
\begin{equation}\label{Drude2}
\partial_{t} {\bf J}({\bf x}',t)+\nu_{e}{\bf J}({\bf x}',t)=\mathcal{N}_e({\bf x}',t){\bf E}({\bf x}',t),
\end{equation}
where $\nu_e$ denotes the effective electron collision frequency. In the MASP model $\mathcal{N}_e({\bf x}',t)$ is computed from
molecular ionization self-consistently:
\begin{equation}\label{El_ND}
\mathcal{N}_e({\bf x}',t) = \mathcal{N}_{e0}({\bf x}')+\mathcal{N}({\bf x}')\sum_{i=1}^m\chi_{\Omega_i}({\bf x}')\mathcal{I}_i(t),
\end{equation}
where the function
\begin{equation}\label{El_ND2}
\mathcal{I}_i(t) = 1-\int_{\Omega_i}|\psi_i({\bf x},t)|^2d^3{\bf x}
\end{equation}
defines a fraction of the electrons freed due to tunnel ionization \cite{keldysh,ammosov}, using the $L^2$-norm of the bound electron wavefunctions.
\begin{rem}\label{rem1}
\Cref{AmpLaws,FarLaws,P_schro,schro,Drude2} form the MASP model.
\end{rem}
\indent In general, for 1-electron TDSE and under Born-Oppenheimer approximation, the MASP model is
3d-3d in the sense, that ${\bf E}({\bf x}',t)$, ${\bf B}({\bf x}',t)$ and $\psi({\bf x},t)$ are functions of spatial coordinates: ${\bf x}'=(x',y',z')^T$ and ${\bf x}=(x,y,z)^T$.  More generally, in the terminology $M$d-$N$d, $M=\dim({\bf x}')$ and $N=\dim({\bf x})$. In practice, we
can reduce the computational complexity of the system by applying a
reasonable dimensionality reduction. For example, \cref{Spa_Dom} illustrates domain decomposition in case of 1d-2d MASP model with the pulse propagating in the $z'$-direction.\\
\indent However, if we want to describe the medium response to a laser pulse within the MASP model accurately, a very large number of TDSEs is required, increasing immensely the overall computational cost of the model. With the purpose to reduce this cost, Lorin et al. \cite{ springer,Lorin2015} have proposed to enrich the MASP model by a simple transport equation for polarization:
\begin{equation}\label{model1_b}
\partial_{t} {\bf P}({\bf x}',t) + v_g\partial_{z'} {\bf P}({\bf x}',t) = {\bf 0},
\end{equation}
where $v_g$ is the group velocity. More specifically, the MEs domain is now decomposed along the $z'$-direction in $N_1$ subdomains, each containing $N_2$ \emph{layers} transversal to $z'$-axis with thickness $\Delta z'$. For the 3d MEs, the dimensionality of such a layer is 2: plane $(x',y')$, while in case of 1d MEs the ``layer'' degenerates to the cell, whose transversal dimension is 0. In order to locally evaluate the dipole moment, the TDSEs are computed only at first layers (or cells) of each subdomain with coordinates denoted ${{\bf x}'}_{\alpha,1}$, where $\alpha=1,\dots,N_1$, and the second index is for the number of layer$/$cell within the subdomain. Thus, we can add to \cref{model1_b} the initial value of polarization ${\bf P}({{\bf x}'}_{\alpha,1},0)=\mathcal{N}({{\bf x}'}_{\alpha,1}){\bf d}({{\bf x}'}_{\alpha,1},0)$. Between these first layers/cells, the evolution equations on ${\bf P}$, such as \cref{model1_b}, are used for a ``cheap''computation (as fully macroscopic) of the polarization vector. For the 1d-2d MASP model this methodology is summarized in \cref{fig0}.\\
\indent Note that this model is applicable as long as the length  $\Delta_{\alpha} z'$ of the subdomains along $z'$-axis is short enough, or$/$and if the molecule density is low enough. That is, as long as the effect of the medium on ${\bf E}$ during the pulse propagation from ${\bf x}'_{\alpha,1}$ to ${\bf x}'_{\alpha+1,1}$ is sufficiently negligible not to be included in the dipole moment calculation of a molecule located at ${\bf x}'_{\alpha+1,1}$. In order to include the medium effects on ${\bf E}$ during the propagation in $({\bf x}'_{\alpha,1},{\bf x}'_{\alpha+1,1})$, an
improvement of the model is necessary.\\
\indent The approach we propose here allows to consider larger propagation lengths and times more accurately than the simple transport equation. Since we consider multidimensional EM field propagation, including forward and backward propagation effects, the equation for polarization is expected to be
of the form:
\begin{equation}\label{GPEE}
\partial_{t}^2{\bf P} -v_g^2\triangle {\bf P} = {\bf S}({\bf E}),
\end{equation}
where for brevity sake we denote ${\bf P}\equiv{\bf P}({\bf x}',t)$, ${\bf E}\equiv{\bf E}({\bf
x}',t)$. The main idea consists of including as much information as possible
to determine ${\bf S}({\bf E})$ in \cref{GPEE}. Cancelling out ${\bf B}$ from \cref{AmpLaws} and \cref{FarLaws}, we get the wave equation for ${\bf E}$:
\begin{equation}\label{wave_for_E}
\partial_{t}^{2} {\bf E} - c^2\triangle {\bf E}+c^2\nabla(\nabla\cdot {\bf E})=-4\pi(\partial_{t}^{2} {\bf P}+\partial_{t} {\bf J}).
\end{equation}
We omit the term $\nabla(\nabla\cdot {\bf E})$, supposing that on the relevant propagation length nonlinearity is not very strong, as $\nabla\cdot{\bf E}\propto\nabla\cdot{\bf P}^{NL}$ \cite{Boyd}:
\begin{equation}\label{N1}
\partial_t^2{\bf E}-c^2\triangle{\bf E}=-4\pi(\partial_t^2{\bf P}+\partial_t{\bf J}).
\end{equation}
Next, we assume that the perturbative expansion
\begin{equation}\label{pertP}
  {\bf P}({\bf x}',t)=\chi^{(1)}{\bf E}({\bf x}',t)+\chi^{(3)}({\bf E}({\bf x}',t)\cdot{\bf E}({\bf x}',t)){\bf E}({\bf x}',t),
\end{equation}
where $\chi^{(1)}$ and $\chi^{(3)}$ are the first and third
\emph{instantaneous} susceptibilities of isotropic medium is an accurate approximation for
${\bf P}$ at least \emph{for short enough propagation length}. From here
\begin{equation}\label{N2}
  {\bf E}({\bf x}',t)=\dfrac{1}{\chi^{(1)}}{\bf P}({\bf x}',t)-\dfrac{\chi^{(3)}}{\chi^{(1)}}({\bf E}({\bf x}',t)\cdot{\bf E}({\bf x}',t)){\bf E}({\bf x}',t).
\end{equation}
Then substituting \cref{N2} in \cref{N1} leads to
\begin{equation}\label{N3}
(1+4\pi\chi^{(1)})\partial_t^2{\bf P}-c^2\triangle{\bf P}=\chi^{(3)}\Big[\partial_t^2\big({\bf E}({\bf E}\cdot{\bf E})\big)-c^2\triangle\big({\bf E}({\bf E}\cdot{\bf E})\big)\Big]-4\pi\chi^{(1)}\partial_t{\bf J},
\end{equation}
where $({\bf E}\cdot{\bf E})=E_x^2+E_y^2+E_z^2$. The group velocity in the first approximation can be defined as
$v_g=c/\sqrt{1+4\pi\chi^{(1)}}$ \cite{Lorin2015}, so that the evolution equation for
${\bf P}$ is shaped into
\begin{equation}\label{eqP2}
\partial_{t}^{2} {\bf P} -v_g^2\triangle {\bf P}=\Big(\dfrac{v_g}{c}\Big)^2\Big\{\chi^{(3)}
\Big[\partial_{t}^{2} ({\bf E}{\bf E}^{2})-c^2\triangle({\bf E}{\bf E}^{2})\Big]- 4\pi\chi^{(1)}\partial_{t}{\bf J}\Big\}.
\end{equation}
At preselected locations, say ${\bf x}'_{\alpha}$ ($\alpha=1,2\dots N_1$) and at each computational
time step $t_{\beta}$, the data ${\bf P}({\bf x}'_{\alpha},t_{\beta})$,
$\partial_t{\bf P}({\bf x}'_{\alpha},t_{\beta})$ are computed from
microscopic TDSEs:
\begin{equation}\label{general_model}
\left\{
\begin{array}{ll}
{\bf P}({\bf x}'_{\alpha},t_{\beta}) = \mathcal{N}({\bf x}'_{\alpha}) \int_{ \mathbb{R}^3}
\big|\psi_{{\bf x}'_{\alpha}}({\bf x},t_{\beta})\big|^2{\bf x}d^3{\bf x}, \\
\partial_t{\bf P}({\bf x}'_{\alpha},t_{\beta}) = \mathcal{N}({\bf x}'_{\alpha})
\int_{\mathbb{R}^3}\partial_t\big|\psi_{{\bf x}'_{\alpha}}({\bf x},t_{\beta})\big|^2{\bf x}d^3{\bf x}.
\end{array}
\right.
\end{equation}
Recall that the current density ${\bf J}$ satisfies the macroscopic kinetic equation \cref{Drude2}. Let the free electron density be a function of the $z'$-coordinate, the direction of the pulse propagation, and time. It is reasonable then to compute this value from the following transport equation:
\begin{equation}\label{free_elec_dens}
  \partial_t \mathcal{N}_e(z',t) + v_g\partial_{z'}\mathcal{N}_e(z',t) = 0,
\end{equation}
with the initial distribution computed according to \cref{El_ND} and \cref{El_ND2}:
\begin{equation}\label{Par_ND}
\mathcal{N}_e(z'_\alpha,t_\beta) = \mathcal{N}_{e0}(z'_\alpha)+\mathcal{N}(z'_\alpha)\sum_{i=1}^m\chi_{\Omega_i}(z'_\alpha)
\Big[1-\int_{\Omega_i}|\psi_i({\bf x},t_\beta)|^2d^3{\bf x}\Big].
\end{equation}
Thus, except for the evaluation of the polarization and the free electron
number density at specific locations, requiring TDSE computations, the evolution equations \cref{eqP2} and \cref{free_elec_dens} are then fully macroscopic. 
\begin{rem}
\Cref{AmpLaws,FarLaws,Drude2} defined on a bounded spatial domain $\Omega$, being coupled to the evolution equations for polarization \cref{eqP2} (or in the simplest case, to equation \cref{model1_b}) and to the evolution equations for free electron density \cref{free_elec_dens} with respective initial conditions \cref{general_model} and \cref{Par_ND}, computed at the set of preselected locations of $\Omega$ through the TDSEs \cref{schro}, form the enriched MASP model.
\end{rem}

\indent It is necessary to comment the philosophy of the proposed approach.
Equations \cref{eqP2} with initial conditions \cref{general_model} form a nonperturbative model,
which includes (i) the transport of  \emph{nonperturbative} data
(as $\partial_t{\bf P}_0$ are computed from TDSEs) from the left boundary to
the right one of each subdomain, and (ii) the \emph{perturbative} corrections
to polarization within these subdomains. In \cref{model1_b}, we do not assume
propagation in a linear medium, however, we use a linear approximation that
is valid when the length of the subdomain is short enough, and the same was assumed while
deriving \cref{eqP2} from \cref{N1} and \cref{pertP}. Thus, in our \emph{model
approach} the LHS in the equation \cref{eqP2} represents the linear transport of the \emph{nonperturbative} polarization $\partial_t{\bf P({{\bf x}'}_\alpha,\cdot)}$ initially taken from the nonperturbative computations \cref{general_model} at reference points ${{\bf x}'}_\alpha$, $\alpha=1,\dots, N_1$, across the subdomains. The RHS of the equation \cref{eqP2} makes the non-zero \emph{perturbative} corrections to the polarization in
interval $\Delta_\alpha z'$. \\
\indent For practical purposes, we simplify the polarization evolution
equation in the case of 1d-2d model, moreover we will consider the case of linearly polarized pulse,
implying that $P_{x'}(z',t)$ as well as $E_{x'}(z',t)$ are null. Thus, assuming that the polarization vector propagates along the $z'$-direction (Laplacian $\triangle$ is reduced to $\partial_{z'}^2$) we obtain an evolution equation
for $P_{y'}$:
\begin{equation}\label{N5}
\begin{array} {l}
  \partial_t^2 P_{y'}-v_g^2\partial_{z'}^2 P_{y'}=3\chi^{(3)}\Big(\dfrac{v_g}{c}\Big)^2E^2_{y'}\Big[\partial_t^2 E_{y'}-c^2\partial_{z'}^2 E_{y'}\Big]+\qquad\qquad\qquad\qquad \\
  \qquad\qquad +6\chi^{(3)}\Big(\dfrac{v_g}{c}\Big)^2E_{y'}\Big[(\partial_t E_{y'})^2-c^2(\partial_{z'} E_{y'})^2\Big]-4\pi\chi^{(1)}\Big(\dfrac{v_g}{c}\Big)^2\partial_tJ_{y'}.
\end{array}
\end{equation}
For the sake of simplicity of the notation we introduce three variable factors
\begin{equation}\label{N6_10}
\begin{array}{l}
a(E_{y'}(z',t)):=1+12\pi\chi^{(3)}(v_g/c)^2E^2_{y'},\\
b(E_{y'}(z',t)):=6\chi^{(3)}(v_g/c)^2E_{y'},\\
g(E_{y'}(z',t)):=4\pi(v_g/c)^2(\chi^{(1)}+3\chi^{(3)}E^2_{y'}),
\end{array}
\end{equation}
and observing that $\partial_t^2 E_{y'}-c^2\partial^2_{z'}
E_{y'}=-4\pi\partial_t^2 P_{y'}-4\pi\partial_t J_{y'}$ in 1d, the polarization wave equation takes the form
\begin{equation}\label{N7}
  \begin{array}{l}
  a(E_{y'})\partial_t^2 P_{y'}-v_g^2\partial_{z'} P_{y'}=\\
  \hspace*{2cm} =b(E_{y'})\Big[(\partial_t E_{y'})^2-c^2(\partial_z E_{y'})^2\Big]-g(E_{y'})\partial_tJ_{y'}.
  \end{array}
\end{equation}
Recall that we want to couple this equation to the MASP model. However, the RHS of \cref{N7} contains partial derivatives of the electric field in time, and this fact may result in some sort of computational instability. To avoid possible instability of the solution, we replace in \cref{N7} $\partial_t
E_{y'}=c\partial_{z'}B_{x'}-4\pi\partial_t P_{y'}-4\pi
J_{y'}$ according to \cref{AmpLaws}, thus
\begin{equation}\label{N7_1}
\begin{array}{l}
   a(E_{y'})\partial_t^2 P_{y'}-v_g^2\partial_{z'}^2 P_{y'}+8\pi b(E_{y'})\Big[(c\partial_{z}B_{x'}-4\pi J_{y'})(\partial_tP_{y'})-2\pi(\partial_tP_{y'})^2\Big] =\\
  =b(E_{y'})\Big\{c^2\Big[(\partial_{z'} B_{x'})^2-(\partial_z E_{y'})^2\Big]+8\pi J_{y'}\Big[2\pi J_{y'}-c\partial_{z'}B_{x'}\Big]\Big\}-g(E_{y'})\partial_tJ_{y'}.
\end{array}
\end{equation}
Finally, we provide the principal details of the procedure leading to the 1d-2d enriched MASP model.
\begin{itemize}
  \item At $t=t_\beta$, all the time derivatives of polarization
      $\partial_t P_{y'}(z'_{\alpha,1},t_\beta)$ are computed at locations with coordinates
      $z'_{\alpha,1}$, $\alpha=1,\dots N_1$, from the set of TDSEs \cref{schro}
      according to \cref{general_model}, see \cref{fig0};
  \item These data are used as boundary conditions for the wave equation
      \cref{N7_1} with $\partial_{z'} B_{x'}(z',t_{\beta})$ and
      $\partial_{z'} E_{y'}(z',t_{\beta})$ in the RHS, in order to compute
      the time derivative of polarization $\partial_t
      P_{y'}(z'_{\alpha,\mu},t_{\beta+1})$ ($\alpha=1,\dots N_1$,
      $\mu=2\dots N_2$) at time $t_{\beta+1}$ and at all other locations
      $\{z'_{\alpha,\mu}\}$ of the MEs domain;
  \item Thus, at time $t_{\beta+1}$ we have $\partial_t
      P_{y'}(z'_{\alpha,1},t_{\beta+1})$, computed using the TDSEs, and
      $\partial_t P_{y'}(z'_{\alpha,\mu},t_{\beta+1})$ $(\mu=2\dots N_2)$,
      computed using the polarization evolution equation. We use these data
      to solve the MEs \cref{AmpLaws,FarLaws} in the domain at time $t_{\beta+1}$.
\end{itemize}
Finally we note that in the case of linearly polarized pulses, it is easy to generalize \cref{N6_10}
including higher order nonlinearities:
\begin{equation}\label{N6_1}
\begin{array}{l}
 a(E_{y'}) :=1+4\pi(v_g/c)^2(3\chi^{(3)}E^2_{y'}+5\chi^{(5)}E^4_{y'}+7\chi^{(7)}E^6_{y'}+\dots), \\
 b(E_{y'}) :=E_{y'}(v_g/c)^2(6\chi^{(3)}+20\chi^{(5)}E_{y'}^2+42\chi^{(7)}E_{y'}^4+\dots),\\
 g(E_{y'}) :=4\pi(v_g/c)^2(\chi^{(1)}+3\chi^{(3)}E^2_{y'}+5\chi^{(5)}E^4_{y'}+7\chi^{(7)}E^6_{y'}+\dots)=\\
  \hspace*{1cm}=4\pi(v_g/c)^2\chi^{(1)}+[a(E_{y'})]^{-1}-1.
\end{array}
\end{equation}

%% file: EandU.tex
\section{Existence and uniqueness of the weak solutions for the MASP models}\label{sec:analysis}
In this section, we study the well-posedness of the MASP models. We start the analysis with the MASP model, see \cref{rem1}, which we will also refer to as the ``pure'' MASP model (in contrast to the ``enriched'' model). We base our proof on the results that were established earlier for the
Schr\"odinger equation \cite{Baud,Iorio,2} and the Maxwell-Schr\"odinger
model \cite{cicp}, which does not include the description of the plasma effects.
Then, we discuss the existence and regularity of solutions
to the model supplemented by the evolution equations for polarization
\cref{GPEE}, and for the free electrons density \cref{free_elec_dens}.
\subsection{Pure MASP Model}
Let $({\bf E}_0, {\bf B}_0, {\bf J}_0,\overline{\psi}_0)^T$ be the
initial data of the equations \cref{AmpLaws,FarLaws,schro,Drude2}, where $\overline{\psi}_0=(\psi_{0,1}\dots\psi_{0,m})^T$, and
$m$ is the total number of cells in the gas region. We suppose that ${\bf
E}_0, {\bf B}_0, {\bf J}_0$ belong to $\big(H^1(\Omega)\big)^3$, where
\begin{itemize}
  \item $\Omega\subset \mathbb{R}^3$ is a bounded spatial domain for
      Maxwell's equations with a smooth boundary
      $\partial\Omega$ where Dirichlet zero BCs be imposedposed for ${\bf E}, {\bf
      B}, {\bf J}$. In other words, we assume that the electromagnetic
      pulse remains always inside the domain $\Omega$, so that for all time the flux of the Poynting vector ${\bf
      S}=\dfrac{c}{4\pi}{\bf E}\times{\bf B}$ (recall, in our case ${\bf
      B}\equiv{\bf H}$) at the boundary $\partial\Omega$ is zero as well;
  \item $H^1(\Omega)$ denotes the Sobolev space $W^{1,2}(\Omega)$ \cite{Taylor}:
\end{itemize}
\begin{equation}\label{Sobol}
H^1(\Omega)=\{v\in L^2(\Omega)\:|\quad\partial_{x_i}v_{\mathcal{D}'}\in L^2(\Omega),\quad i=1,2,3\},
\end{equation}
where the derivative $\partial_{x_i}v_{\mathcal{D}'}$ is defined in the
distributional (weak) sense, and $L^p$ $(1\leq
p<+\infty)$ denote the Lebesgue spaces. Further in this section, we will omit the symbol $\mathcal{D}'$ on the
derivatives, for the sake of simplicity of the notation. \\
\indent In the Born-Oppenheimer approximation the distance
between nuclei $R_0>0$ is a fixed parameter of the wavefunctions. We suppose that
$\overline{\psi}_0\in\big(H^1(\mathbb{R}^3)\cap H_1(\mathbb{R}^3)\big)^m$,
where the norm of the Sobolev space $H^1(\mathbb{R}^3)$ reads
\begin{equation}\label{Sobolev_norm}
  \|\psi\|^2_{H^1}=\displaystyle\int_{\mathbb{R}^3}\big(|\psi({\bf x},R_0)|^2+
  \displaystyle\sum_{i=1}^3|\partial_{x_i}\psi({\bf x},R_0)|^2\big)d{\bf x}.
\end{equation}
The norm of \emph{another} Sobolev space $H_1$ (with \emph{low} index), which
is the image of $H^1$ (with \emph{upper} index) under the Fourier transform
\cite{Baud,Taylor}, is defined as:
\begin{equation}\label{FT_Sobolev}
    \|\psi\|^2_{H_1}=\displaystyle\int_{\mathbb{R}^3}(1+|({\bf x},{\bf R}_0)|^2)
  |\psi({\bf x},R_0)|^2d{\bf x},
\end{equation}
where $|({\bf x},{\bf R}_0)|^2:=|{\bf x}-{\bf R}_0|^2=|{\bf x}-{\bf
e}_0R_0|^2$ with unit vector ${\bf e}_0$ that defines the orientation of the
molecule in plane $(x,y)$.\\
\indent Now we extend the Existence and Uniqueness Theorem \cite{cicp},
referred to solutions of the Maxwell-Shr\"odinger (MS) model, to the case of
MASP model.
\begin{theorem}\label{EUMaxScro}
Suppose that $({\bf E}_0, {\bf B}_0, {\bf
J}_0)\in\big(H^1(\Omega)\big)^3\times\big(H^1(\Omega)\big)^3\times\big(H^1(\Omega)\big)^3$,
$\overline{\psi}_0\in(H^1\cap H_1)^m$, $R_0>0$ is a constant for all $t\in
\mathbb{R}_+$ and $\mathcal{N}\in C_0^\infty(\Omega)$. Suppose that on the
smooth boundary $\partial\Omega$ zero Dirichlet BCs are imposed on vectors
${\bf E}, {\bf B}, {\bf J}$ for all $t\in\mathbb{R}_+$. Then there exists a
time $T>0$, and a unique solution $({\bf E}, {\bf B}, {\bf J},
\overline{\psi})\in \big(L^\infty(0,T;(H^1(\Omega))^3)\cap
H^1(0,T;(L^2(\Omega))^3)\big)^3\times L^\infty(0,T;(H^1\cap H_1)^m)$ to
Equations \cref{AmpLaws,FarLaws,schro,Drude2}.
\end{theorem}
In order to prove this theorem, we should prove several important
intermediate results. The first lemma follows closely \cite{cicp} and
\cite{Baud}.
\begin{lemma}\label{LemB_2}
Suppose that ${\bf E}({\bf x}',\cdot)\in L^\infty(0,T)$ and $\partial_t{\bf
E}({\bf x}',\cdot)\in L^1(0,T)$ for ${\bf x}'$ fixed in $\Omega$. We assume
that $R_0>0$ is fixed for all $t\in \mathbb{R}_+$. Then for all
$\psi_{0,i}\in H^1\cap H_1$, there exists $\psi_i\in L^\infty(0,T;H^1\cap
H_1)$ solution to the Schr\"odinger equation \cref{schro}, and there exists
a constant $C_T>0$ such that
\begin{equation*}
  \|\psi_i\|_{L^\infty(0,T;H^1\cap H_1)}\leq C_T\|\psi_{0,i}\|_{H^1\cap H_1}.
\end{equation*}
\end{lemma}
\noindent The proof can be found in \cite{cicp}. \\
\indent The next lemma from \cite{cicp} generalizes the result of
\cref{LemB_2}, on the vector-valued function
$\overline{\psi}=(\psi_1\dots\psi_m)^T$:
\begin{lemma}[Lorin et al. \cite{cicp}]\label{LemB_3}
Suppose given ${\bf E}({\bf x}',\cdot)\!\in L^\infty(0,T)$ and $\partial_t{\bf
E}({\bf x}',\cdot)\!\in L^1(0,T)$, for ${\bf x}'$ fixed in $\Omega$. Then there
exists $C_T>0$, such that for all $\overline{\psi}_0\in (H^1\cap H_1)^m$
there exists a solution $\overline{\psi}\in L^\infty(0,T;(H^1\cap H_1)^m)$
and
\begin{equation*}
  \|\overline{\psi}\|_{L^\infty(0,T;(H^1\cap H_1)^m)}\leq C_T\|\overline{\psi}_0\|_{(H^1\cap H_1)^m}.
\end{equation*}
\end{lemma}
\begin{proof} It follows from the previous lemma. \end{proof}
\begin{lemma}[Lorin et al.\cite{cicp}]\label{LemB_4}
For all $T>0$, and $\overline{\psi}\in L^\infty(0,T;(H^1\cap H_1)^m)$, ${\bf
P}\in L^\infty(0, T; (\mathcal{C}_0^\infty)^3)$.
\end{lemma}
\begin{proof} Recall that $\chi_{\Omega_i}$, $\mathcal{N}$ belong
to $\mathcal{C}^\infty_0(\Omega)$ and $\psi_i\in L^\infty(0,T;H^1\cap H_1)$,
for all $i=1,\dots,m$, we then deduce that
\begin{equation*}
  t \mapsto\int_{\mathbb{R}^3\times\mathbb{R}_+}\psi_i^*(R_0,{\bf x},t){\bf x}\psi_i(R_0,{\bf x},t)\in L^\infty(0,T).
\end{equation*}
In particular for all $i=1,\dots,m$, as $\psi_i$ belongs to $H^1\cap H_1$, and the integral is defined. Finally, by definition of polarization \cref{P_schro}:\\
\indent ${\bf P}({\bf x}',t) = -\mathcal{N}({\bf x}')\displaystyle\sum_{i=1}^m\chi_{\Omega_i}({\bf x}')\displaystyle\int_{R^3}\psi^*_i(R_0,{\bf x},t){\bf x}\psi_i(R_0,{\bf x},t)d{\bf x}dR_0$,\\
which concludes the proof. \end{proof}
\begin{lemma}[Lorin et al.\cite{cicp}]\label{LemB_5}
For ${\bf x}'$ fixed in $\Omega$ and $T>0$, $\partial_t{\bf P}({\bf
x}',\cdot)\in L^\infty(0,T)$ and $\partial_t(\nabla\cdot{\bf P}({\bf
x}',\cdot))\in L^\infty(0,T)$.
\end{lemma}
\begin{proof} First, we observe that $\partial_t{\bf P}({\bf
x}',t)=\mathcal{N}({\bf x}')\displaystyle\sum_{i=1}^m \partial_t{\bf
P}_i({\bf x}',t)$, where
\begin{equation*}
\begin{array}{l}
  \partial_t{\bf P}_i({\bf x}',t)=\chi_{\Omega_i}({\bf x}')\displaystyle\int_{R^3}\partial_t\psi^*_i(R_0,{\bf x},t){\bf x}\psi_i(R_0,{\bf x},t)d{\bf x}dR_0+\\
  \qquad\qquad +\chi_{\Omega_i}({\bf x}')\displaystyle\int_{R^3}\psi^*_i(R_0,{\bf x},t){\bf x}\partial_t\psi_i(R_0,{\bf x},t)d{\bf x}dR_0.
\end{array}
\end{equation*}
As $\psi_i\in L^\infty(0,T;H^1\cap H_1)$ and $\chi_{\Omega_i}$,
 $\mathcal{N}\in\mathcal{C}^\infty_0(\Omega)$ then $\partial_t{\bf P}_i\in
L^\infty (0,T)$ for all $i=1,\dots,m$, hence $\partial_t{\bf P}\in L^\infty
(0,T)$. Also as, according to \cref{LemB_4}, $\nabla\cdot{\bf
P}(\cdot,t)\in \mathcal{C}^\infty_0(\Omega)$ at $t$ fixed, we also have that
$\partial_t(\nabla\cdot{\bf P})({\bf x}',\cdot)\in L^\infty (0,T).$ \end{proof}
\begin{lemma}\label{LemB_6}
Suppose given $\mathcal{N}_{e0}\in \mathcal{C}_0^\infty$, then for all $T>0$,
and $\overline{\psi}\in L^\infty(0,T;(H^1\cap H_1)^m)$, $\mathcal{N}_e\in
L^\infty(0,T;\mathcal{C}_0^\infty)$.
\end{lemma}
\begin{proof} We know that $\mathcal{N}$ and $\chi_{\Omega_i}$
being to $\mathcal{C}_0^\infty$. Following Equations \cref{El_ND},
\cref{El_ND2}, the expression for density of free electron
$\mathcal{N}_e({\bf x}',t)$ at different ${\bf x}'$ reads
\begin{equation*}
  \mathcal{N}_e({\bf x}',t)=\mathcal{N}_{e0}({\bf x}')+
  \mathcal{N}({\bf x}')\sum_{i=1}^m\chi_{\Omega_i}({\bf x}')
  \Big(1-\int_{\Omega_i}|\psi_i(R_0,{\bf x}',t)|^2d{\bf x}\Big)dR_0,
\end{equation*}
which proves the lemma. \end{proof}
\begin{lemma}\label{LemB_1}
Suppose that ${\bf E}(t)|_{\partial\Omega}={\bf B}(t)|_{\partial\Omega}={\bf
0}\in \mathbb{R}^3$, then for all $T>0$ and $\Omega\subset\mathbb{R}^3$
\begin{equation}\label{LE1}
  \begin{array}{l}
     \|{\bf E}(\cdot,T)\|^2_{\big(L^2(\Omega)\big)^3}+\|{\bf B}(\cdot,T)\|^2_{\big(L^2(\Omega)\big)^3}=
     \|{\bf E}_0(\cdot)\|^2_{\big(L^2(\Omega)\big)^3}+\|{\bf B}_0(\cdot)\|^2_{\big(L^2(\Omega)\big)^3}-\\
      -8\pi\displaystyle\int_0^T\int_\Omega{\bf E}({\bf x}',t)\cdot\partial_t{\bf P}({\bf x}',t)d{\bf x}'dt
      -8\pi\displaystyle\int_0^T\int_\Omega{\bf E}({\bf x}',t)\cdot{\bf J}({\bf x}',t)d{\bf x}'dt,
  \end{array}
\end{equation}
and
\begin{equation}\label{LE2}
  \begin{array}{l}
     \|\nabla\!\cdot\!{\bf E}(\cdot,T)\|^2_{L^2(\Omega)}=
     \|\nabla\!\cdot\!{\bf E}_0(\cdot)\|^2_{L^2(\Omega)}
     -8\pi\displaystyle\int_0^T\int_\Omega\nabla\!\cdot\!{\bf E}({\bf x}',t)\partial_t\nabla\!\cdot\!{\bf P}({\bf x}',t)d{\bf x}'dt-\\
     \qquad\qquad\qquad\qquad\qquad\qquad -
     8\pi\displaystyle\int_0^T\int_\Omega\nabla\!\cdot\!{\bf E}({\bf x}',t)\nabla\!\cdot\!{\bf J}({\bf x}',t)d{\bf x}'dt.
  \end{array}
\end{equation}
\end{lemma}
\begin{proof} We respectively use Ampere-Maxwell's \cref{AmpLaws} and
Faraday-Maxwell's \cref{FarLaws} laws:
We take the scalar products of \cref{AmpLaws} with ${\bf E}$, and
\cref{FarLaws} with ${\bf B}$ that result in
\begin{equation}\label{LE0}
    \begin{array}{l}
  {\bf E}({\bf x}',t)\cdot\partial_t{\bf E}({\bf x}',t)=c{\bf E}({\bf x}',t)\cdot[\nabla\times{\bf B}({\bf x}',t)]-\\
  \qquad\qquad\qquad\qquad\qquad-4\pi{\bf E}({\bf x}',t)\cdot\partial_t{\bf P}({\bf x}',t)-4\pi{\bf E}({\bf x}',t)\cdot{\bf J}({\bf x}',t),\\
  {\bf B}({\bf x}',t)\cdot\partial_t{\bf B}({\bf x}',t)=-c{\bf B}({\bf x}',t)\cdot[\nabla\times{\bf E}({\bf x}',t)].
    \end{array}
\end{equation}
Noticing that in the LHSs ${\bf E}({\bf x}',t)\cdot\partial_t{\bf E}({\bf
x}',t)=\partial_t{\bf E}^2({\bf x}',t)/2$ and ${\bf B}({\bf
x}',t)\cdot\partial_t{\bf B}({\bf x}',t)=\partial_t{\bf B}^2({\bf
x}',t)/2$, so that we obtain after integration over $(0,T)\times\Omega$:
\begin{align*}
  \int_0^T\int_\Omega{\bf E}({\bf x}',t) \cdot\partial_t{\bf E}({\bf x}',t)d{\bf x}'dt=
  \dfrac{1}{2}\big(\|{\bf E}(\cdot,T)\|^2_{\big(L^2(\Omega)\big)^3}-\|{\bf E}_0(\cdot)\|^2_{\big(L^2(\Omega)\big)^3}\big),\\
  \int_0^T\int_\Omega{\bf B}({\bf x}',t) \cdot\partial_t{\bf B}({\bf x}',t)d{\bf x}'dt=
  \dfrac{1}{2}\big(\|{\bf B}(\cdot,T)\|^2_{\big(L^2(\Omega)\big)^3}-\|{\bf B}_0(\cdot)\|^2_{\big(L^2(\Omega)\big)^3}\big).
\end{align*}
We add the expressions \cref{LE0} using that $\nabla\cdot[{\bf E}\times{\bf B}]={\bf B}\cdot[\nabla\times{\bf E}]-{\bf E}\cdot[\nabla\times{\bf B}]$, and integrating over $(0,T)\times\Omega$. Finally, using the divergence
theorem, we deduce \cref{LE1}.\\
\indent Now we apply the operator $\nabla$ to both sides of
\cref{AmpLaws}. Since the divergence of the curl is 0, we obtain the
equation
\begin{equation}\label{LE3}
  \partial_t\nabla\cdot{\bf E}({\bf x}',t)=-4\pi\partial_t\nabla\cdot{\bf P}({\bf x}',t)-4\pi\nabla\cdot{\bf J}({\bf x}',t).
\end{equation}
We take the product of the scalar equation \cref{LE3} with a scalar
$\nabla\cdot{\bf E}$ and again integrate over $(0,T)\times\Omega$ resulting in \cref{LE2}. \end{proof}
\begin{lemma}\label{LemB_9}
Suppose that ${\bf E}(t)|_{\partial\Omega}={\bf B}(t)|_{\partial\Omega}={\bf
0}\in \mathbb{R}^3$, then for all $T>0$ and $\Omega\subset\mathbb{R}^3$
\begin{equation}\label{LE1_1}
  \begin{array}{l}
     \|\partial_t{\bf E}(\cdot,T)\|^2_{\big(L^2(\Omega)\big)^3}+c^2\|\nabla\times{\bf E}(\cdot,T)\|^2_{\big(L^2(\Omega)\big)^3}=
     \|\partial_t{\bf E}_0(\cdot)\|^2_{\big(L^2(\Omega)\big)^3}+\\ \qquad+\|\nabla\times{\bf E}_0(\cdot)\|^2_{\big(L^2(\Omega)\big)^3}-
      8\pi\displaystyle\int_0^T\int_\Omega\partial_t{\bf E}({\bf x}',t)\cdot\partial_t^2{\bf P}({\bf x}',t)d{\bf x}'dt-\\
      \qquad\qquad-8\pi\displaystyle\int_0^T\int_\Omega\partial_t{\bf E}({\bf x}',t)\cdot\partial_t{\bf J}({\bf x}',t)d{\bf x}'dt,
  \end{array}
\end{equation}
and
\begin{equation}\label{LE2_1}
  \begin{array}{l}
     \|\partial_t{\bf B}(\cdot,T)\|^2_{\big(L^2(\Omega)\big)^3}+c^2\|\nabla\times{\bf B}(\cdot,T)\|^2_{\big(L^2(\Omega)\big)^3}=
     \|\partial_t{\bf B}_0(\cdot)\|^2_{\big(L^2(\Omega)\big)^3}+\\ \qquad+\|\nabla\times{\bf B}_0(\cdot)\|^2_{\big(L^2(\Omega)\big)^3}+
      8\pi c\displaystyle\int_0^T\int_\Omega\partial_t{\bf B}({\bf x}',t)\cdot\partial_t[\nabla\times{\bf P}({\bf x}',t)]d{\bf x}'dt+\\
      \qquad\qquad+8\pi c\displaystyle\int_0^T\int_\Omega\partial_t{\bf B}({\bf x}',t)\cdot[\nabla\times{\bf J}({\bf x}',t)]d{\bf x}'dt.
  \end{array}
\end{equation}
\end{lemma}
\begin{proof} We differentiate \cref{AmpLaws} in time and then
multiply both sides by $\partial_t{\bf E}$. We note that
\begin{align*}
  &\partial_t{\bf E}\cdot\partial_t^2{\bf E} = \dfrac{1}{2}\partial_t(\partial_t{\bf E}\cdot\partial_t{\bf E})=\dfrac{1}{2}\partial_t|\partial_t{\bf E}|^2, \\
  &\partial_t{\bf E}\cdot[\nabla\times[\nabla\times{\bf E}]] = \nabla\cdot[[\nabla\times{\bf E}]\times\partial_t{\bf E}]+[\nabla\times{\bf E}]\cdot[\nabla\times\partial_t{\bf E}]= \\
  &\qquad\qquad\qquad= \nabla\cdot[[\nabla\times{\bf E}]\times\partial_t{\bf E}]+\dfrac{1}{2}\partial_t|\nabla\times{\bf E}|^2.
\end{align*}
We integrate the equation over $(0,T)\times\Omega$, apply the divergence
theorem and obtain \cref{LE1_1}. Similarly, by differentiation
\cref{FarLaws} in time, then by multiplication by $\partial_t{\bf B}$ and
finally by integration over $(0,T)\times\Omega$ we come to \cref{LE2_1}.
\end{proof}
\begin{lemma}\label{LemB_7}
For all $T>0$,
\begin{equation}\label{LE4}
    \begin{array}{l}
  \|{\bf J}(\cdot,T)\|^2_{\big(L^2(\Omega)\big)^3}=\|{\bf J}_0(\cdot)\|^2_{\big(L^2(\Omega)\big)^3}-
   2\nu_e\displaystyle\int_0^T\|{\bf J}(\cdot,t)\|^2_{\big(L^2(\Omega)\big)^3}dt+\\
   \qquad\qquad+2\displaystyle\int_0^T\int_\Omega\mathcal{N}_e({\bf x}',t){\bf J}({\bf x}',t)\cdot{\bf E}({\bf x}',t)d{\bf x}'dt
   \end{array}
\end{equation}
and
\begin{equation}\label{LE4_2}
    \begin{array}{l}
  \|\partial_t{\bf J}(\cdot,T)\|^2_{\big(L^2(\Omega)\big)^3}=\|\partial_t{\bf J}_0(\cdot)\|^2_{\big(L^2(\Omega)\big)^3}-
   2\nu_e\displaystyle\int_0^T\|\partial_t{\bf J}(\cdot,t)\|^2_{\big(L^2(\Omega)\big)^3}dt+\\
   \qquad\qquad+2\displaystyle\int_0^T\int_\Omega\mathcal{N}_e({\bf x}',t)\partial_t{\bf J}({\bf x}',t)\cdot\partial_t{\bf E}({\bf x}',t)d{\bf x}'dt+ \\
   \qquad\qquad\qquad\qquad+2\displaystyle\int_0^T\int_\Omega\partial_t\mathcal{N}_e({\bf x}',t)\partial_t{\bf J}({\bf x}',t)\cdot{\bf E}({\bf x}',t)d{\bf x}'dt
   \end{array}
\end{equation}
and
\begin{equation}\label{LE5}
    \begin{array}{l}
  \|{\nabla\cdot\bf J}(\cdot,T)\|^2_{L^2(\Omega)}=\|{\nabla\cdot\bf J}_0(\cdot)\|^2_{L^2(\Omega)}-
   2\nu_e\displaystyle\int_0^T\|{(\nabla\cdot\bf J})(\cdot,t)\|^2_{L^2(\Omega)}dt+\\
   \qquad\qquad+2\displaystyle\int_0^T\int_\Omega\nabla\mathcal{N}_e({\bf x}',t)\cdot{\bf E}({\bf x}',t)\nabla\cdot{\bf J}({\bf x}',t)d{\bf x}'dt+\\
   \qquad\qquad\qquad+2\displaystyle\int_0^T\displaystyle\int_\Omega\mathcal{N}_e({\bf x}',t)\nabla\cdot{\bf E}({\bf x}',t)\nabla\cdot{\bf J}({\bf x}',t)d{\bf x}'dt
   \end{array}
\end{equation}
and finally
\begin{equation}\label{LE5_2}
    \begin{array}{l}
  \|{\nabla\times\bf J}(\cdot,T)\|^2_{\big(L^2(\Omega)\big)^3}=\|{\nabla\times\bf J}_0(\cdot)\|^2_{\big(L^2(\Omega)\big)^3}-
   2\nu_e\displaystyle\int_0^T\|{(\nabla\times\bf J})(\cdot,t)\|^2_{\big(L^2(\Omega)\big)^3}dt+\\
   \qquad\qquad+2\displaystyle\int_0^T\int_\Omega[\nabla\mathcal{N}_e({\bf x}',t)\times{\bf E}({\bf x}',t)]\cdot[\nabla\times{\bf J}({\bf x}',t)]d{\bf x}'dt+\\
   \qquad\qquad\qquad+2\displaystyle\int_0^T\displaystyle\int_\Omega\mathcal{N}_e({\bf x}',t)[\nabla\times{\bf E}({\bf x}',t)]\cdot[\nabla\times{\bf J}({\bf x}',t)]d{\bf x}'dt.
   \end{array}
\end{equation}
\end{lemma}
\begin{proof} We take the scalar product of \cref{Drude2} with ${\bf J}$
\begin{equation*}
\dfrac{1}{2}\partial_{t} |{\bf J}({\bf x}',t)|^2=-\nu_{e}|{\bf J}({\bf x}',t)|^2+\mathcal{N}_e({\bf x}',t){\bf J}({\bf x}',t)\cdot{\bf E}({\bf x}',t).
\end{equation*}
Integrating over $(0,T)\time\Omega$, gives \cref{LE4}. To prove
\cref{LE4_2}, we differentiate Equation \cref{Drude2} in time, then we take
the scalar product of both parts with $\partial_t{\bf J}$,  and integrate over $(0,T)\time\Omega$.\\
\indent By applying $\nabla\cdot$ on \cref{Drude2}, and then integrating
over $(0,T)\time\Omega$ with $\nabla\cdot{\bf J}$, we deduce \cref{LE5}.
Finally, by applying $\nabla\times$ on \cref{Drude2}, and then integrating
over $(0,T)\time\Omega$ with $\nabla\times{\bf J}$, we deduce \cref{LE5_2}.
\end{proof}
\begin{rem}\label{PhysDim} With a view of proper management of the physical variable dimensions in
the following lemmas, we should
 make a remark. Recall that we use atomic units for all the equations, so that from MEs \cref{AmpLaws,FarLaws}
 we can conclude that $\dim({\bf E})=\mathrm{T}\dim({\bf J})$, where $\mathrm{T}$ is the symbol for the time dimension.
 Taking into account Drude's model \cref{Drude2}, we can see that $\dim(\mathcal{N}_e)=\mathrm{T}^{-2}$.
 In the following, we introduce a constant $\eta>0$, having the dimension of time, $\dim(\eta)=\mathrm{T}$.
 \end{rem}
 \begin{lemma}\label{LemB_8}
There exists a constant $C>0$ such that for all time $T>0$
\begin{equation}\label{LB10}
  \sup_{0\leq t\leq T}\|{\bf E}(\cdot,t)\|^2_{(H^1(\Omega))^3}+\sup_{0\leq t\leq T}\|{\bf B}(\cdot,t)\|^2_{(H^1(\Omega))^3}
  +\eta^2\sup_{0\leq t\leq T}\|{\bf J}(\cdot,t)\|^2_{(H^1(\Omega))^3}\leq C.
\end{equation}
\end{lemma}
\begin{proof} From the result \cref{LE1} of \cref{LemB_1} we
have for all $t\in(0,T]$
\begin{equation*}
  \begin{array}{l}
     \|{\bf E}(\cdot,T)\|^2_{\big(L^2(\Omega)\big)^3}+\|{\bf B}(\cdot,T)\|^2_{\big(L^2(\Omega)\big)^3}\leq
     \|{\bf E}_0(\cdot)\|^2_{\big(L^2(\Omega)\big)^3}+\|{\bf B}_0(\cdot)\|^2_{\big(L^2(\Omega)\big)^3}+\\
      +8\pi\displaystyle\int_0^T\int_\Omega|{\bf E}({\bf x}',t)\cdot\partial_t{\bf P}({\bf x}',t)|d{\bf x}'dt+
      8\pi\int_0^T\displaystyle\int_\Omega|{\bf E}({\bf x}',t)\cdot{\bf J}({\bf x}',t)|d{\bf x}'dt\leq\\
      \leq \|{\bf E}_0(\cdot)\|^2_{\big(L^2(\Omega)\big)^3}+\|{\bf B}_0(\cdot)\|^2_{\big(L^2(\Omega)\big)^3}+
      8\pi\eta^{-1}\displaystyle\int_0^T \|{\bf E}(\cdot,t)\|^2_{\big(L^2(\Omega)\big)^3}dt+\\
      4\pi\eta\displaystyle\int_0^T\|\partial_t{\bf P}(\cdot,t)\|^2_{\big(L^2(\Omega)\big)^3}dt
      +4\pi\eta\displaystyle\int_0^T\|{\bf J}(\cdot,t)\|^2_{\big(L^2(\Omega)\big)^3}dt. \\
      \qquad\qquad
  \end{array}
\end{equation*}
Similarly from \cref{LE2}:
\begin{equation*}
  \begin{array}{l}
     \|\nabla\!\cdot\!{\bf E}(\cdot,T)\|^2_{L^2(\Omega)}\leq
     \|\nabla\!\cdot\!{\bf E}_0(\cdot)\|^2_{L^2(\Omega)}+8\pi\displaystyle\int_0^T\int_\Omega|\nabla\!\cdot\!{\bf E}({\bf x}',t)\partial_t\nabla\!\cdot\!{\bf P}({\bf x}',t)|d{\bf x}'dt+\\
     \qquad+8\pi\displaystyle\int_0^T\int_\Omega|\nabla\!\cdot\!{\bf E}({\bf x}',t)\nabla\!\cdot\!{\bf J}({\bf x}',t)|d{\bf x}'dt\leq
      \|\nabla\!\cdot\!{\bf E}_0(\cdot)\|^2_{L^2(\Omega)}+\\
      \qquad\qquad+8\pi\eta^{-1}\displaystyle\int_0^T\|\nabla\!\cdot\!{\bf E}(\cdot,t)\|^2_{L^2(\Omega)}dt+4\pi\eta\displaystyle\int_0^T\|\partial_t\nabla\!\cdot\!{\bf P}(\cdot,t)\|^2_{L^2(\Omega)}dt+\\
      \qquad\qquad\qquad\qquad +4\pi\eta\displaystyle\int_0^T\|\nabla\!\cdot\!{\bf J}(\cdot,t)\|^2_{L^2(\Omega)}dt.\\
       \qquad\qquad
  \end{array}
\end{equation*}
From \cref{LemB_6} and \cref{LE4} of
\cref{LemB_7}, we have for all $t\in(0,T]$
\begin{equation*}
    \begin{array}{l}
  \|{\bf J}(\cdot,T)\|^2_{\big(L^2(\Omega)\big)^3}\leq\|{\bf J}_0(\cdot)\|^2_{\big(L^2(\Omega)\big)^3}+
   2\nu_e\displaystyle\int_0^T\|{\bf J}(\cdot,t)\|^2_{\big(L^2(\Omega)\big)^3}dt+\qquad\qquad\\
   \qquad\qquad+2\displaystyle\int_0^T\int_\Omega\mathcal{N}_e({\bf x}',t)|{\bf J}({\bf x}',t)\cdot{\bf E}({\bf x}',t)|d{\bf x}'dt\leq\|{\bf J}_0(\cdot)\|^2_{\big(L^2(\Omega)\big)^3}+\\
   +(2\nu_e+\eta^{-1})\displaystyle\int_0^T\|{\bf J}(\cdot,t)\|^2_{\big(L^2(\Omega)\big)^3}dt+\eta{\displaystyle\sup_{\Omega,[0,T]}}\mathcal{N}_e^2({\bf x}',t)\displaystyle\int_0^T\|{\bf E}(\cdot,t)\|^2_{\big(L^2(\Omega)\big)^3}dt,
   \end{array}
\end{equation*}
\noindent and with \cref{LemB_6} and \cref{LE5}
\begin{equation*}
    \begin{array}{l}
  \|{\nabla\cdot\bf J}(\cdot,T)\|^2_{L^2(\Omega)}\leq\|{\nabla\cdot\bf J}_0(\cdot)\|^2_{L^2(\Omega)}+
   2\nu_e\int_0^T\displaystyle\int_\Omega{|(\nabla\cdot\bf J})({\bf x}',t)|^2d{\bf x}'dt+\\
   \qquad+2\int_0^T\displaystyle\int_\Omega|\nabla\mathcal{N}_e({\bf x}',t)\cdot{\bf E}({\bf x}',t)\nabla\cdot{\bf J}({\bf x}',t)|d{\bf x}'dt+\\
   \qquad\qquad+2\int_0^T\displaystyle\int_\Omega|\mathcal{N}_e({\bf x}',t)\nabla\cdot{\bf E}({\bf x}',t)\nabla\cdot{\bf J}({\bf x}',t)|d{\bf x}'dt\leq\|{\nabla\cdot\bf J}_0(\cdot)\|^2_{L^2(\Omega)}+\\
   \qquad\qquad
   +2(\nu_e+\eta^{-1})\displaystyle\int_0^T\|(\nabla\cdot{\bf J})(\cdot,t)\|^2_{L^2(\Omega)}dt+\\
   \qquad\qquad\qquad+\eta{\displaystyle\sup_{\Omega,[0,T]}}(\nabla\mathcal{N}_e({\bf x}',t))^2\displaystyle\int_0^T\|{\bf E}(\cdot,t)\|^2_{L^2(\Omega)}dt+\\
   \qquad\qquad\qquad\qquad+\eta{\displaystyle\sup_{\Omega,[0,T]}}\mathcal{N}_e^2({\bf x}',t)\displaystyle\int_0^T\|\nabla\cdot{\bf E}(\cdot,t)\|^2_{L^2(\Omega)}d{\bf x}'dt.
   \end{array}
\end{equation*}
\noindent Combining the first and third inequalities, we get:
\begin{equation}\label{LE12}
  \begin{array}{l}
     \|{\bf E}(\cdot,T)\|^2_{\big(L^2(\Omega)\big)^3}+\|{\bf B}(\cdot,T)\|^2_{\big(L^2(\Omega)\big)^3}+
     \eta^2\|{\bf J}(\cdot,T)\|^2_{\big(L^2(\Omega)\big)^3}\leq\qquad\\
      \leq\|{\bf E}_0(\cdot)\|^2_{\big(L^2(\Omega)\big)^3}+\|{\bf B}_0(\cdot)\|^2_{\big(L^2(\Omega)\big)^3}+
      \eta^2\|{\bf J}_0(\cdot)\|^2_{\big(L^2(\Omega)\big)^3}+\\
      \qquad+(8\pi\eta^{-1}+\eta^3{\displaystyle\sup_{\Omega,[0,T]}}\mathcal{N}_e^2({\bf x}',t) )\displaystyle\int_0^T\|{\bf E}(\cdot,t)\|^2_{\big(L^2(\Omega)\big)^3}dt+\\
      +(4\pi\eta+2\nu_e\eta^2+\eta)\displaystyle\int_0^T\|{\bf J}(\cdot,t)\|^2_{\big(L^2(\Omega)\big)^3}dt+4\pi\eta\displaystyle\int_0^T\|\partial_t{\bf P}(\cdot,t)\|^2_{\big(L^2(\Omega)\big)^3}dt.
  \end{array}
\end{equation}
Identically for the second and fourth inequalities:
\begin{equation}\label{LE21}
  \begin{array}{l}
     \|\nabla\!\cdot\!{\bf E}(\cdot,T)\|^2_{L^2(\Omega)}+\eta^2\|\nabla\!\cdot\!{\bf J}(\cdot,T)\|^2_{L^2(\Omega)}\leq     \|\nabla\!\cdot\!{\bf E}_0(\cdot)\|^2_{L^2(\Omega)}+\eta^2\|\nabla\!\cdot\!{\bf J}_0(\cdot)\|^2_{L^2(\Omega)}+\\
     \qquad\qquad+(8\pi\eta^{-1}+\eta^3{\displaystyle\sup_{\Omega,[0,T]}}\mathcal{N}_e^2({\bf x}',t))\displaystyle\int_0^T\|\nabla\!\cdot\!{\bf E}(\cdot,t)\|^2_{L^2(\Omega)}dt+\\
     \qquad\qquad+(4\pi\eta+2\nu_e\eta^2+2\eta)\displaystyle\int_0^T\|\nabla\!\cdot\!{\bf J}(\cdot,t)\|^2_{L^2(\Omega)}dt+\\
     +\eta^3{\displaystyle\sup_{\Omega,[0,T]}}(\nabla\mathcal{N}_e({\bf x}',t))^2\displaystyle\int_0^T\|{\bf E}(\cdot,t)\|^2_{L^2(\Omega)}dt+4\pi\eta\displaystyle\int_0^T\|\partial_t\nabla\!\cdot\!{\bf P}(\cdot,t)\|^2_{L^2(\Omega)}dt.
  \end{array}
\end{equation}
In addition to \cref{LE21}, we recall that $\nabla\cdot{\bf B}({\bf
x}',t)\equiv0$. According to \cref{LemB_5,LemB_6},
$\partial_t{\bf P}\in L^\infty(0, T; (\mathcal{C}_0^\infty)^3)$ and
$\mathcal{N}_e({\bf x}',t)\in L^\infty(0,T;\mathcal{C}_0^\infty)$, so using
Gr\"onwall's inequality we can confirm boundedness of the norms in the LHS of inequalities \cref{LE12,LE21}.\\
\indent Similarly, using \cref{LemB_9} and equations \cref{LE4_2,LE5_2}, we can conclude boundedness of the following combination of the
norms:
\begin{align*}
  \|\partial_t{\bf E}(\cdot,T)\|^2_{\big(L^2(\Omega)\big)^3}+c^2\|\nabla\times{\bf E}(\cdot,T)\|^2_{\big(L^2(\Omega)\big)^3}+
  \|\partial_t{\bf B}(\cdot,T)\|^2_{\big(L^2(\Omega)\big)^3}+\qquad\qquad\qquad
  \\+c^2\|\nabla\times{\bf B}(\cdot,T)\|^2_{\big(L^2(\Omega)\big)^3}+
  \eta^2\Big\{\|\partial_t{\bf J}(\cdot,T)\|^2_{\big(L^2(\Omega)\big)^3}+c^2\|\nabla\times{\bf J}(\cdot,T)\|^2_{\big(L^2(\Omega)\big)^3}\Big\}.
\end{align*}
From here we deduce \cref{LB10}. \end{proof}
\begin{proof} of \cref{EUMaxScro}. So far, we have proven
that for all $T>0$, there exists a constant $C>0$ such that:
\begin{equation*}
    \begin{array}{r}
  \|{\bf E}\|^2_{L^\infty\big(0,T;(H^1(\Omega))^3\big)\cap H^1\big(0,T;(L^2(\Omega))^3\big)}+
  \|{\bf B}\|^2_{L^\infty\big(0,T;(H^1(\Omega))^3\big)\cap H^1\big(0,T;(L^2(\Omega))^3\big)}+\\
  \qquad\qquad+\eta^2\|{\bf J}\|^2_{L^\infty\big(0,T;(H^1(\Omega))^3\big)\cap H^1\big(0,T;(L^2(\Omega))^3\big)}+\mu\|\overline{\psi}\|_{L^\infty\big(0,T;(H^1\cap H_1)^m\big)}\leq C,
    \end{array}
\end{equation*}
where $\mu>0$ is a dimensional constant, such that $\dim(\mu)=\dim({\bf
E}^2)$. The boundness of the last term in the above inequality is a
consequence of \cref{LemB_2}. Now as
$L^\infty\big(0,T;(H^1(\Omega))^3\big)\times L^\infty\big(0,T;(H^1\cap
H_1)^m\big)$ is compactly embedded in $L^2(\Omega\times(0,T])\times
(L^2(\mathbb{R}^3))^m$ by Leray-Schauder's fixed point theorem, we deduce the
existence of a solution for \cref{AmpLaws,FarLaws,schro,Drude2}. The approach is the same as described in \cite{Yin}. It is based on the introduction of a continuous mapping derived from
\cref{AmpLaws,FarLaws,schro}, which depends on a parameter
$\lambda\in[0,1]$ and admits a fixed point in
$L^2(\Omega\times(0,T])\times (L^2(\mathbb{R}^3))^m$ as verifying Leray-Schauder's theorem assumptions \cite{Owen}.\\
\indent \emph{Uniqueness} is proven by a classical process of introducing two
solutions of the Cauchy problem \cref{AmpLaws,FarLaws,schro}:
$({\bf E}_1,{\bf B}_1,{\bf J}_1,\overline{\psi}_1)^T$ and $({\bf E}_2,{\bf
B}_2,{\bf J}_2,\overline{\psi}_2)^T$, and denoting their difference as $({\bf
E},{\bf B},{\bf J},\overline{\psi})^T:=({\bf E}_2-{\bf E}_1,{\bf B}_2-{\bf
B}_1,{\bf J}_2-{\bf J}_1,\overline{\psi}_2-\overline{\psi}_1)^T$ with ICs
$\big({\bf E}(\cdot,0),{\bf B}(\cdot,0),{\bf J}(\cdot,0),
\overline{\psi}(\cdot,0)\big)^T=({\bf 0},{\bf 0},{\bf 0},{\bf 0})^T$. As in
\cref{LemB_1,LemB_2}, see also for details \cite{cicp}, there
exists a constant $C>0$ such that:
\begin{equation}\label{uniqC}
  \begin{array}{c}
  \dfrac{d}{dt}\Big\{\mu\|\overline{\psi}(t)\|^2_{(H^1\cap H_1)^m}+\displaystyle\int_{\Omega}\big(\|{\bf E}({\bf x}',t)\|^2_{\big(L^2(\Omega)\big)^3}+\qquad\qquad\qquad\qquad\\
  +\|{\bf B}({\bf x}',t)\|^2_{\big(L^2(\Omega)\big)^3}+
  \eta^2\|{\bf J}({\bf x}',t)\|^2_{\big(L^2(\Omega)\big)^3}\big)d{\bf x}'\Big\}\leq\qquad\qquad\\
  \qquad\qquad C\Big\{\mu\|\overline{\psi}(t)\|^2_{(H^1\cap H_1)^m}+\displaystyle\int_{\Omega}\big(\|{\bf E}({\bf x}',t)\|^2_{\big(L^2(\Omega)\big)^3}+\\
  \qquad\qquad\qquad\qquad+\|{\bf B}({\bf x}',t)\|^2_{\big(L^2(\Omega)\big)^3}+
  \eta^2\|{\bf J}({\bf x}',t)\|^2_{\big(L^2(\Omega)\big)^3}\big)d{\bf x}'\Big\}.
  \end{array}
\end{equation}
Using Gr\"onwall's inequality we conclude $\big({\bf E}(\cdot,t),{\bf
B}(\cdot,t),
{\bf J}(\cdot,t),\overline{\psi}(\cdot,t)\big)^T\equiv({\bf 0},{\bf 0},{\bf 0},{\bf 0})^T$ for all $t>0$. Moreover, the polarization is also unique, ${\bf P}={\bf P}_2-{\bf
P}_1\equiv{\bf 0}$ for $t\geq0$, as by definition:
\begin{equation*}
    \begin{array}{l}
  {\bf P}({\bf x}',t)=\mathcal{N}({\bf x}')\displaystyle\sum_{i=1}^m{\bf P}_i({\bf x}',t)=\\
  \qquad=\mathcal{N}({\bf x}')\displaystyle\sum_{i=1}^m\chi_{\Omega_i}({\bf x}')\displaystyle\int_{\mathbb{R}^3\times \mathbb{R}_+}{\bf x}
  \big(|\psi_{i,2}(R_0,{\bf x},t)|^2-|\psi_{i,1}(R_0,{\bf x},t)|^2\big)d{\bf x}dR_0,
    \end{array}
\end{equation*}
and as $\overline{\psi}(\cdot,t)\equiv{\bf 0}$ for all $t>0$.
\end{proof}

\indent The main practical consequence of \cref{EUMaxScro} in the
sense of numerical treatment of the model is the conservation in time of
regularity of the solutions, which justifies the use of finite difference
methods (FDM) for solving equations formed the MASP model \cite{2,cicp}. 
In addition, we note that the FDM is a natural choice as the geometry of 
the computational domains of the MEs and TDSEs are simple enough.
\subsection{MASP Model Supplemented by Evolution Equations for Polarization and Free Electron Density}
This part discusses the issue, whether we could still rely on the same
regularity of solutions in case of the MASP model enriched by the evolution
equations for polarization and free electrons density. Certainly, it is too
complicated to prove the theorem of existence and uniqueness for the MASP
model with embedded general evolution equation \cref{N7} which, moreover, is
nonlinear. However, since we intend to use this equation in the regime of
relatively weak nonlinearity (with purpose to obtain a correction to the
solution of homogeneous equation), it makes sense to consider at least the
MASP model complemented by the homogeneous wave equations for polarization
vector :
\begin{equation}\label{WE_P}
\partial_{t}^2{\bf P} -v_g^2\triangle{\bf P} = 0,
\end{equation}
with velocity $v_g$, considered as a constant. Note that if we apply the
operator $\nabla$ to both sides of \cref{WE_P} and recall
that the divergence of the polarization vector equals to the density of the
bound charges taken with opposite sign, $\nabla\cdot{\bf
P}=-(\mathcal{N}-\mathcal{N}_e)$, we obtain the evolution equation for the
free electron density:
\begin{equation}\label{WE_Ne}
\partial_{t}^2\mathcal{N}_e -v_g^2\triangle\mathcal{N}_e = 0,
\end{equation}
while the afore-mentioned equation \cref{free_elec_dens} describes
one-directional wave only. To obtain \cref{WE_Ne}, we recall that the gas
density depends on spatial coordinate only, $\mathcal{N}=\mathcal{N}({\bf
x}')$, and we make an assumption that this dependence is close to constant,
i.e. $\nabla\mathcal{N}({\bf x}')={\bf 0}$ and $\triangle\mathcal{N}({\bf x}')=0$ for all ${\bf x}'\in \Omega$. \\
\indent We intend to solve the system of equations
\cref{AmpLaws,FarLaws,schro,Drude2,WE_P,WE_Ne} subject to the initial data set $({\bf E}_0, {\bf B}_0,{\bf J}_0, {\psi}_{0}, {\bf P}_0, \partial_t{\bf P}_0, \mathcal{N}_{e0}, \partial_t \mathcal{N}_{e0})^T$. For simplicity, we assume that the bounded domain of  Maxwell's equations $\Omega$
contains only \emph{one} ``subdomain'' $\widetilde{\Omega}$ with a gas, which means that now ${\psi}_{0}\equiv\psi_{0,1}$. Using these microscopic data, we can compute ${\bf
P}(t)|_{\partial\widetilde{\Omega}}$, see \cref{P_schro}, and
$\mathcal{N}(t)|_{\partial\widetilde{\Omega}}$ see \cref{El_ND},
\cref{El_ND2}, and their time derivatives in the first cell of the gas
region, i.e. on the boundary of the gas region. We suppose that initially
\emph{within the domain}, the polarization vector and free electron number
density equal zero: ${\bf P}_0={\bf 0}\in \mathbb{R}^3$, $\partial_t{\bf
P}_0={\bf 0}\in \mathbb{R}^3$, $\mathcal{N}_{e0}=0\in\mathbb{R}$, $\partial_t
\mathcal{N}_{e0}=0\in\mathbb{R}$. This provides for wave equations
\cref{WE_P,WE_Ne} the initial-boundary value problems. Now we
state the existence theorem for the supplemented MASP model.
\begin{theorem}\label{EUMaxScroEvol}
Suppose that $({\bf E}_0, {\bf B}_0, {\bf J}_0)\in\big(H^1(\Omega)\big)^3\times\big(H^1(\Omega)\big)^3\times\big(H^1(\Omega)\big)^3$, ${\psi}_0\in(H^1\cap H_1)$, $R_0>0$ is a constant for all $t\in \mathbb{R}_+$ and $\mathcal{N}\in C_0^\infty(\Omega)$. Suppose that on the smooth boundary $\partial\Omega$ zero Dirichlet BCs are imposed for vectors ${\bf E}, {\bf B}, {\bf J}$, ${\bf P}$ for all $t\in\mathbb{R}_+$. In addition, on the smooth boundary $\partial\widetilde{\Omega}$ of the gas region the values of ${\bf P}(t)\Big|_{\partial\widetilde{\Omega}}$, $\mathcal{N}_{e}(t)\Big|_{\partial\widetilde{\Omega}}$ are computed via ${\psi}(R_0,{\bf x},t)$ from \cref{schro,El_ND,El_ND2} with $m=1$, while for $t=0$ $({\bf P}_0, \partial_t{\bf P}_0, \mathcal{N}_{e0}, \partial_t\mathcal{N}_{e0})=({\bf 0},{\bf 0}, 0, 0)$. Then
there exists a time $T>0$, and a unique solution $({\bf E}, {\bf B}, {\bf J},{\bf P}, {\psi})\in \big(L^\infty(0,T;(H^1(\Omega))^3)\cap H^1(0,T;(L^2(\Omega))^3)\big)^4\times L^\infty(0,T;(H^1\cap H_1))$ to Equations \cref{AmpLaws,FarLaws,schro} (with $m=1$),
\cref{Drude2,WE_P,WE_Ne}.
\end{theorem}
In the case of the MASP model supplemented by the evolution equations
\cref{WE_P,WE_Ne}, we first need to adjust several lemmas proven
above. In particular, \cref{LemB_4,LemB_5} are still true for ${\bf x}'={{\bf x}'}_1$, i.e. in the first gas ``cell'' only. However, inside the gas subdomain we have to consider the evolution equations, in particular
\begin{equation}\label{DE1}
  \partial_{t}^2{\bf P} =v_g^2\triangle {\bf P}.
\end{equation}
\begin{lemma}\label{LemD_1}
We assume Dirichlet BCs on $\partial\Omega$: ${\bf
P}(t)|_{\partial\Omega}={\bf 0}\in \mathbb{R}^3$, then for all $T>0$, and
$\Omega\subset\mathbb{R}^3$
\begin{equation}\label{DE5}
\begin{array}{l}
\|\partial_t{\bf P}(\cdot,T)\|^2_{\big(L^2(\Omega)\big)^3}+v_g^2\|\nabla\times{\bf P}(\cdot,T)\|^2_{\big(L^2(\Omega)\big)^3}=\|(\partial_t{\bf P})_0(\cdot)\|^2_{\big(L^2(\Omega)\big)^3}+\\
+v_g^2\|(\nabla\times{\bf P})_0(\cdot)\|^2_{\big(L^2(\Omega)\big)^3}+2v_g^2\int_0^T\int_\Omega \partial_t{\bf P}({\bf x}',t)\cdot \nabla \mathcal{N}_e({\bf x}',t)d{\bf x}'dt.
\end{array}
\end{equation}
\end{lemma}
\begin{proof} We multiply \cref{DE1} by $\partial_{t}{\bf P}$ and
consider separately the LHS and the RHS:
\begin{align}\label{DE2}
  \partial_{t}{\bf P}\cdot\partial_{t}^2{\bf P}& =\dfrac{1}{2}\partial_t|\partial_t{\bf P}|^2,\\
  \label{DE3}
  \partial_{t}{\bf P}\cdot\triangle{\bf P}&=(\partial_{t}{\bf P}\cdot\nabla)(\nabla\cdot{\bf P})-\partial_{t}{\bf P}\cdot[\nabla\times[\nabla\times{\bf P}]],
\end{align}
where $|\partial_t{\bf P}|^2={\partial_t\bf P}\cdot{\partial_t\bf P}$. \\
\indent Recall that $\nabla\cdot{\bf P}=-(\mathcal{N}-\mathcal{N}_e)$, so the
first term of the RHS reads
\begin{center}
  $(\partial_t{\bf P}\cdot\nabla)(\nabla\cdot{\bf P})=-(\partial_t{\bf P}\cdot\nabla)(\mathcal{N}-\mathcal{N}_e)=\partial_t{\bf P}\cdot \nabla \mathcal{N}_e.$
\end{center}
The second term in \cref{DE3} can be rewritten
\begin{center}
  $\partial_t{\bf P}\cdot[\nabla\times[\nabla\times{\bf P}]] = \nabla\cdot[[\nabla\times{\bf P}]\times\partial_t{\bf P}]+[\nabla\times{\bf P}]\cdot[\nabla\times\partial_t{\bf P}]=$ \\
  $\qquad\qquad\qquad\qquad= \nabla\cdot[[\nabla\times{\bf P}]\times\partial_t{\bf P}]+\dfrac{1}{2}\partial_t|\nabla\times{\bf P}|^2$,
\end{center}
so finally we come to
\begin{equation}\label{DE4}
  \partial_t\big\{|\partial_t{\bf P}|^2+v_g^2|\nabla\times{\bf P}|^2\big\}=2v_g^2\partial_t{\bf P}\cdot \nabla \mathcal{N}_e-2v_g^2\nabla\cdot[[\nabla\times{\bf P}]\times\partial_t{\bf P}].
\end{equation}
Now we integrate \cref{DE4} over $(0,T)\times\Omega$ (applying the divergence theorem and using the BCs) to obtain \cref{DE5}. \end{proof}

\indent Similarly, in addition to \cref{LemB_6} we need to consider the property
of $\mathcal{N}_e$ inside the subdomain.
\begin{lemma}\label{LemD_2}
We assume Dirichlet BCs on $\partial\Omega$:
$\mathcal{N}_e(t)|_{\partial\Omega}=0\in \mathbb{R}$, then for all $T>0$ and
$\Omega\subset\mathbb{R}^3$
\begin{equation}\label{DE6}
\begin{array}{l}
 \|\partial_t\mathcal{N}_e(\cdot,T)\|^2_{L^2(\Omega)}+v_g^2\|\nabla\mathcal{N}_e(\cdot,T)\|^2_{L^2(\Omega)}=\\
 \hspace*{3cm}=\|(\partial_t \mathcal{N}_e)_0(\cdot)\|^2_{L^2(\Omega)}+v_g^2\|(\nabla \mathcal{N}_e)_0(\cdot)\|^2_{L^2(\Omega)}.
\end{array}
\end{equation}
\end{lemma}
\begin{proof} We multiply the wave equation \cref{WE_Ne} by
$\partial_t \mathcal{N}_e$ and note that
\begin{align}
  \partial_t\mathcal{N}_e\partial_t^2\mathcal{N}_e &= \dfrac{1}{2}(\partial_t \mathcal{N}_e)^2 \\
  \nonumber \partial_t\mathcal{N}_e\triangle\mathcal{N}_e &= \nabla\cdot(\partial_t\mathcal{N}_e\nabla\mathcal{N}_e)-(\nabla\mathcal{N}_e)\cdot(\nabla\partial_t\mathcal{N}_e)=\\
  &=\nabla\cdot(\partial_t\mathcal{N}_e\nabla\mathcal{N}_e)-\dfrac{1}{2}\partial_t(\nabla\mathcal{N}_e)^2.
\end{align}
Then we integrate both sides over $(0,T)\times\Omega$ to $\cref{DE6}$.
\end{proof}
\begin{lemma}\label{LemD_3}
There exists a constant $C_1>0$, such that for all $T>0$
\begin{equation}\label{DE111}
  \sup_{0\leq t\leq T}\|\mathcal{N}_e(\cdot,t)\|^2_{H^1(\Omega)}\leq C_1.
\end{equation}
\end{lemma}
\begin{proof} From
$\partial_t\mathcal{N}_e^2=2\mathcal{N}_e\partial_t\mathcal{N}_e$ we deduce
\begin{align}\label{DE112}
  \nonumber \|\mathcal{N}_e(\cdot,T)\|^2_{L^2(\Omega)} \leq \|\mathcal{N}_{e0}\|^2_{L^2(\Omega)}&+\eta^{-1}\int_0^T\|\mathcal{N}_e(\cdot,T)\|^2_{L^2(\Omega)}dt+\\
   &+ \eta\int_0^T\|\partial_t\mathcal{N}_e(\cdot,T)\|^2_{L^2(\Omega)}dt.
\end{align}
From now on in current section, $\eta$ is defined in the same way as in
\cref{PhysDim}. Then from \cref{DE112,DE6} and Gr\"onwall's inequality we come to \cref{DE111}. \end{proof}
\begin{lemma}\label{LemD_4}
There exists a constant $C_2>0$ such that for all $T>0$
\begin{equation}\label{DE11}
  \sup_{0\leq t\leq T}\|{\bf P}(\cdot,t)\|^2_{(H^1(\Omega))^3}\leq C_2.
\end{equation}
\end{lemma}
\begin{proof} From \cref{LemD_1} and for all $t>0$
\begin{equation}\label{DE8}
\begin{array}{r}
 \|\partial_t{\bf P}(\cdot,T)\|^2_{\big(L^2(\Omega)\big)^3}+v_g^2\|\nabla\times{\bf P}(\cdot,T)\|^2_{\big(L^2(\Omega)\big)^3}\leq\|(\partial_t{\bf P})_0\|^2_{\big(L^2(\Omega)\big)^3}+\\
+v_g^2\|(\nabla\times{\bf P})_0\|^2_{\big(L^2(\Omega)\big)^3}+\eta^{-1}\displaystyle\int_0^T \|\partial_t{\bf P}(\cdot,t)\|^2_{\big(L^2(\Omega)\big)^3}dt+\\
+\eta v_g^4\displaystyle\int_0^T\|\nabla\mathcal{N}_e(\cdot,t)\|^2_{L^2(\Omega)}dt.
\end{array}
\end{equation}
Thus, we can conclude by Gr\"onwall's inequality and \cref{DE111} that
$\|\partial_t{\bf P}(\cdot,T)\|^2_{\big(L^2(\Omega)\big)^3}$ and
$\|\nabla\times{\bf P}(\cdot,T)\|^2_{\big(L^2(\Omega)\big)^3}$ are bounded.
We recall that according \cref{LemD_3} and relation $\nabla\cdot{\bf
P}=-(\mathcal{N}-\mathcal{N}_e)$, norm $\|\nabla\cdot{\bf
P}(\cdot,T)\|^2_{L^2(\Omega)}$ is bounded as well. Hence from $\partial_t{\bf
P}^2={\bf P}\partial_t{\bf P}$ we obtain
\begin{align*}
  \|{\bf P}(\cdot,T)\|^2_{\big(L^2(\Omega)\big)^3}\leq \|{\bf P}_0\|^2_{\big(L^2(\Omega)\big)^3}+\qquad\qquad\qquad\qquad\qquad\qquad\qquad\\
  \qquad+\eta^{-1}\displaystyle\int_0^T\|{\bf P}(\cdot,t)\|^2_{\big(L^2(\Omega)\big)^3}dt+\eta\displaystyle\int_0^T\|\partial_t{\bf P}(\cdot,t)\|^2_{\big(L^2(\Omega)\big)^3}dt,
\end{align*}
thus, again with Gr\"onwall's inequality and \cref{DE8} we deduce \cref{DE11}. \end{proof}

\indent Note that from \cref{LemD_3} we deduce boundedness of
$\|\partial_t(\nabla\cdot{\bf P})(\cdot,T)\|^2_{L^2(\Omega)}=
\|\partial_t\mathcal{N}_e(\cdot,T)\|^2_{L^2(\Omega)}$ and then deduce
\cref{DE11}. That means, all needed conditions to prove \cref{LemB_8} are
satisfied. Now we can extend \cref{LemB_8} in the case of the
enriched MASP model.
\begin{lemma}\label{LemD_5}
There exists a constant $C>0$ such that for all $T>0$
\begin{equation}\label{DE7}
\begin{array}{l}
  \displaystyle\sup_{0\leq t\leq T}\|{\bf E}(\cdot,t)\|^2_{(H^1(\Omega))^3}+\displaystyle\sup_{0\leq t\leq T}\|{\bf B}(\cdot,t)\|^2_{(H^1(\Omega))^3}+\\
  \qquad+\eta^2\displaystyle\sup_{0\leq t\leq T}\|{\bf J}(\cdot,t)\|^2_{(H^1(\Omega))^3}
  +\displaystyle\sup_{0\leq t\leq T}\|{\bf P}(\cdot,t)\|^2_{(H^1(\Omega))^3}\leq C.
\end{array}
\end{equation}
\end{lemma}
\begin{proof}
We can deduce this result from \cref{LemB_8} and \cref{LemD_4}.\end{proof}
\begin{proof} of \cref{EUMaxScroEvol}. As in the proof of
\cref{EUMaxScro}, the existence is based on Leray-Schauder's fixed
point theorem \cite{Owen}. In \cref{LemB_1,LemB_2,LemB_3,LemB_4,LemB_5,LemB_6,LemB_7,LemB_8,LemB_9} and
\cref{LemD_1,LemD_2,LemD_3,LemD_4}, we proved that for all $T>0$ there exists a
constant $C>0$ such that:
\begin{equation*}
    \begin{array}{l}
  \|{\bf E}\|^2_{L^\infty\big(0,T;(H^1(\Omega))^3\big)\cap H^1\big(0,T;(L^2(\Omega))^3\big)}+
  \|{\bf B}\|^2_{L^\infty\big(0,T;(H^1(\Omega))^3\big)\cap H^1\big(0,T;(L^2(\Omega))^3\big)}+\\
  \hspace*{0.1cm}+\eta^2\|{\bf J}\|^2_{L^\infty\big(0,T;(H^1(\Omega))^3\big)\cap H^1\big(0,T;(L^2(\Omega))^3\big)}+\|{\bf P}\|^2_{L^\infty\big(0,T;(H^1(\Omega))^3\big)\cap H^1\big(0,T;(L^2(\Omega))^3\big)}+\\
  \hspace*{4cm}+\mu\|{\psi}\|_{L^\infty\big(0,T;(H^1\cap H_1)\big)}\leq C,
    \end{array}
\end{equation*}
and $L^\infty\big(0,T;(H^1(\Omega))^3\big)\times L^\infty\big(0,T;(H^1\cap
H_1)\big)$ is compactly embedded in $L^2(\Omega\times(0,T])\times
(L^2(\mathbb{R}^3\times \mathbb{R}_+))$. The approach follows \cite{Yin}. We
introduce a continuous mapping derived from \cref{AmpLaws,FarLaws,schro,Drude2,WE_P,WE_Ne}, 
that depends on a parameter $\lambda\in[0,1]$ and that admits a fixed point in
$L^2(\Omega\times(0,T])\times (L^2(\mathbb{R}^3\times\mathbb{R}_+))$ as verifying Leray-Schauder's theorem assumptions.\\
\indent To prove uniqueness, we take the difference
vector $({\bf E},{\bf B},{\bf J},{\bf P},\overline{\psi})^T:=({\bf E}_2-{\bf
E}_1,{\bf B}_2-{\bf B}_1,{\bf J}_2-{\bf J}_1,{\bf P}_2-{\bf
P}_1,\overline{\psi}_2-\overline{\psi}_1)^T$, with zero ICs, where 
$({\bf E}_1,{\bf B}_1,{\bf J}_1,{\bf P}_1,\overline{\psi}_1)^T$
and $({\bf E}_2,{\bf B}_2,{\bf J}_2,{\bf P}_2,\overline{\psi}_2)^T$ denote
two solutions. Then applying the methods presented in the above lemmas, see
also \cite{cicp}, we come to the inequality with some constant $C>0$:
\begin{equation}\label{uniqC2}
  \begin{array}{c}
  \dfrac{d}{dt}\Big\{\mu\|\overline{\psi}(t)\|^2_{(H^1\cap H_1)}+\displaystyle\int_{\Omega}\big(\|{\bf E}({\bf x}',t)\|^2_{\big(L^2(\Omega)\big)^3}+\|{\bf B}({\bf x}',t)\|^2_{\big(L^2(\Omega)\big)^3}\qquad\qquad\\
  +\eta^2\|{\bf J}({\bf x}',t)\|^2_{\big(L^2(\Omega)\big)^3}+\|{\bf P}({\bf x}',t)\|^2_{\big(L^2(\Omega)\big)^3}\big)d{\bf x}'\Big\}\leq\qquad\\
  \qquad C\Big\{\mu\|\overline{\psi}(t)\|^2_{(H^1\cap H_1)}+\displaystyle\int_{\Omega}\big(\|{\bf E}({\bf x}',t)\|^2_{\big(L^2(\Omega)\big)^3}+\|{\bf B}({\bf x}',t)\|^2_{\big(L^2(\Omega)\big)^3}+\\
  \qquad\qquad+\eta^2\|{\bf J}({\bf x}',t)\|^2_{\big(L^2(\Omega)\big)^3}+\|{\bf P}({\bf x}',t)\|^2_{\big(L^2(\Omega)\big)^3}\big)d{\bf x}'\Big\}.
  \end{array}
\end{equation}
We conclude by Gr\"onwall's inequality that $\big({\bf E}(\cdot,t),{\bf
B}(\cdot,t),{\bf J}(\cdot,t),{\bf
P}(\cdot,t),\overline{\psi}(\cdot,t)\big)^T\equiv({\bf 0},{\bf 0},{\bf
0},{\bf 0})^T$ for all $t>0$. \end{proof}
\indent Summarizing the outcomes of this part, because of the regularity of the solutions not only of the ``pure'' MASP model, but the ``extended'' MASP model as well.

%% file: parallel.tex
\section{Numerical Methods and Parallel Computing}\label{sec:numerics}
\begin{figure}[b!]
\center{\includegraphics[scale=0.82, trim=50 25 0 15, clip]{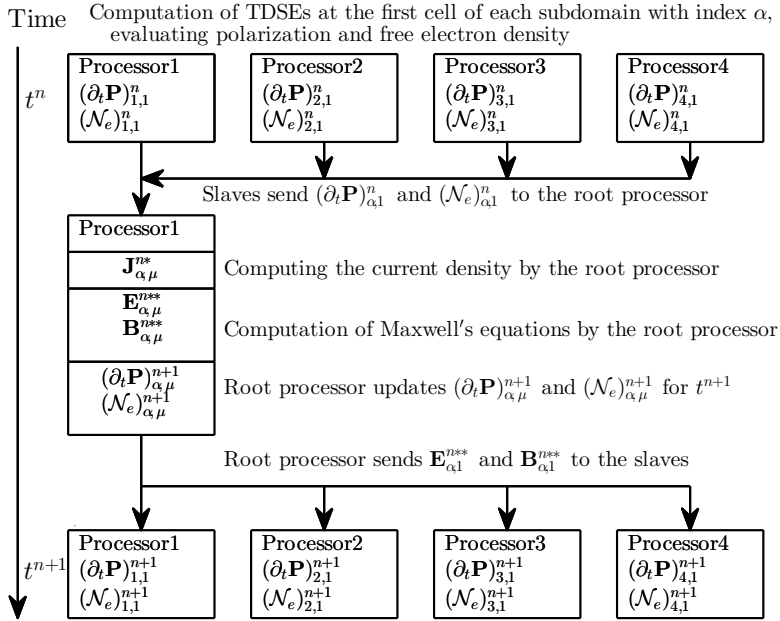}}
\caption[Parallelization: 1d-2d MASP model with the polarization equation.]{Example of parallelization: 1d-2d MASP model enriched by the evolution equations.
Index of subdomain $\alpha=1\dots N_1$ (here $N_1=4$), index of the cell in the subdomain $\mu=1\dots N_2$. Length of the gas region $L=4\times N_2\times\Delta z'$.}
\label{parall_1d_2}
\end{figure}

For 1d Maxwell’s equations we use the second order Lax-Wendroff scheme, see e.g. \cite{Strik}, which is stable under the Courant-Fredrichs-Lewy (CFL)-condition. We solve the 2d Schr\"odinger equation in 3 stages using the symmetric Strang splitting of second order \cite{strang}. At the second stage, we solve the equation with static Coulomb potential using Crank-Nicolson's scheme, which is unconditionally stable. Note that we deal with the multiscale problem, where the time step $\Delta t_S$ for Schr\"odinger's equation is much smaller than for Maxwell's equations $\Delta t_M$: $\Delta t_M/\Delta t_S=10\dots 20$, see also \cite{Lyt}. Finally, to preserve the second order of convergence of the overall scheme, we use Adams-Bashforth's second order consistent method to solve Drude's equation \cref{Drude2} and Lax-Wendroff-type scheme for solving the evolution equation for polarization in its simplest \cref{model1_b} or nonlinear \cref{N7_1} form.\\
\indent Now we discuss the coupling of the schemes approximating (i) the MASP
model, (ii) the polarization evolution equation \cref{eqP2}, and (iii) the free electron density evolution equation \cref{free_elec_dens}. As an example, we assume that the gas region $L$, is divided in 4 subdomains each containing $N_2$ Maxwell's cells, see \cref{fig0}.
At time $t^n$ each of 4 processors solves the TDSE only in the first cell of the subdomain assigned to it,
i.e. at nodes $z'_{\alpha,1}$, where $\alpha=\{1,2,3,4\}$ is the ordinal number of the subdomain, and the second index corresponds to the position number within the subdomain, see \cref{parall_1d_2}.
Then the data $(\partial_t{\bf P})_{\alpha,1}^n$ and $(\mathcal{N}_e)_{\alpha,1}^n$ are sent to the root processor in order to update (i)
the current of free electrons ${\bf J}_{\alpha,\mu}^{n*}$, and then (ii) the EM field at time $t^{n**}$: ${\bf E}_{\alpha,\mu}^{n**}$,
${\bf B}_{\alpha,\mu}^{n**}$, $\mu=1,\dots, N_2$.
Note that at this point, we assume that the root processor knows the values of $(\partial_t{\bf P})_{\alpha,\mu}^n$ and
$(\mathcal{N}_e)_{\alpha,\mu}^n$ for \emph{all} spatial points ${z'}_{\alpha,\mu}$ ($\alpha=1,\dots,4$, $\mu=1,\dots, N_2$), but not only for the first cells
$\{z'_{\alpha, 1}\}$ where the TDSEs are solved.
To clarify that apparent contradiction, we point out that at the time $t\in(t^{n**},t^{n+1})$, the root processor evaluates $(\partial_t {\bf P})_{\alpha,\mu}^{n+1}$
and $(\mathcal{N}_e)_{\alpha,\mu}^{n+1}$ for the \emph{next} time cycle, in all Maxwell's
cells using the evolution equations for the polarization and the free
electron density. Finally, we generalize the described scheme in the case of $m$ subdomains
and $l$ processors, requiring the ratio $m/l$ to be an integer.

%% file: results.tex
\section{Numerical Experiments}\label{sec:experiments}
In this section, we present some results obtained with the enriched MASP model, and compare them with those obtained with the ``pure'' MASP model.
\subsection{The MASP Model Enriched by the Simple Evolution Equation}
\begin{figure}[b!]
\center{\includegraphics[scale=0.85]{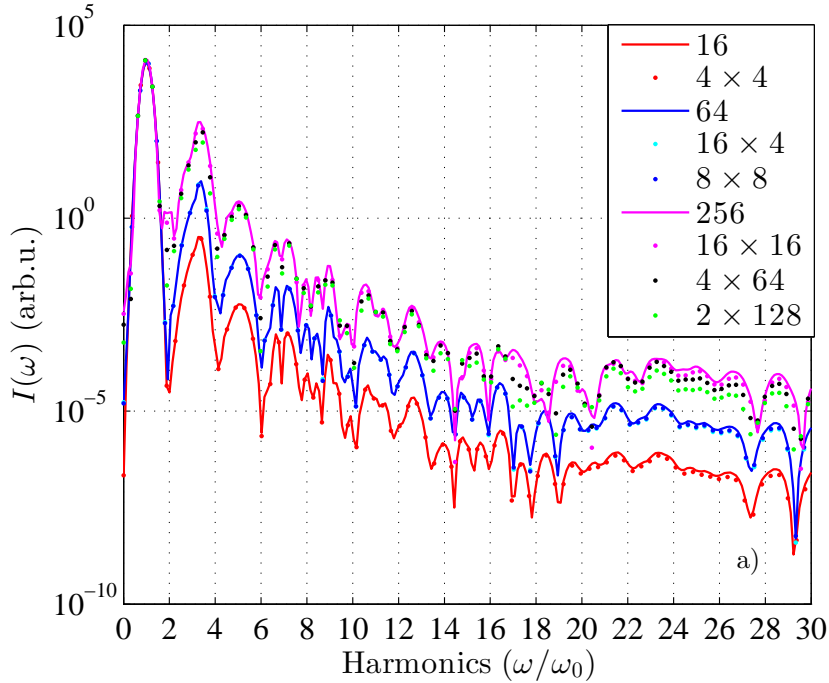}}
\caption{Spectral intensities of the electric field harmonics for different gas region lengths;
$\lambda=800$ nm, $I=5\times10^{13}$ $\mathrm{W}/\mathrm{cm}^2$,
$\Delta z_M=100$ a.u., so that $L=100\times N$ a.u. (see the legend for values of $N=N_1\times N_2$).
The grid for solving 2-d TDSE is $300\times 300$ with step $\Delta x_S=\Delta y_S=0.3$ a.u.,
gas density $\mathcal{N}=5.17\times 10^{-5}$ a.u.}
\label{LP_PE}
\end{figure}

\begin{figure}[t!]
\center{\includegraphics[scale=0.75]{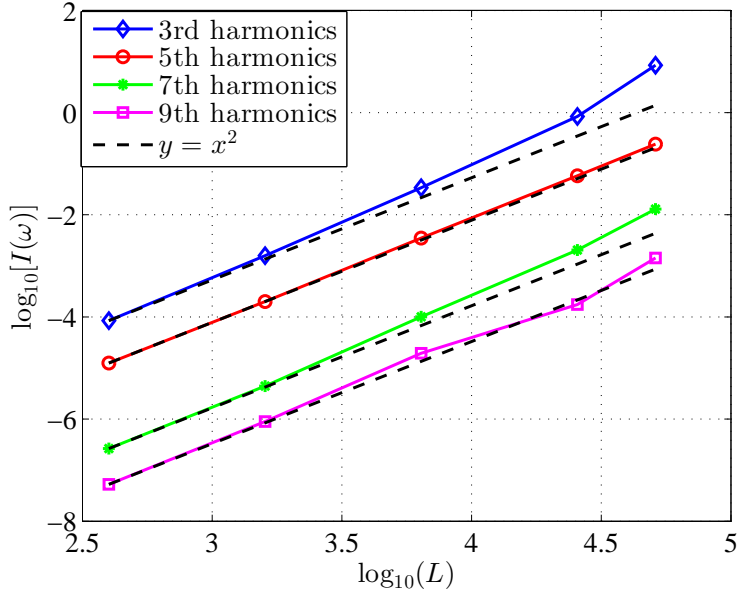}}
\caption{Intensity of the low order harmonics of the circularly polarized pulse as a function of the propagation length in a gas: $L=100\times N$ $(N=4,16,64,16\times 16\:and\:32\times 16)$ a.u.; $\lambda=800$ nm, $I=5\times10^{13}$ $\mathrm{W}/\mathrm{cm}^2$. The grid for solving 2-d TDSE is $500\times 500$ with step $\Delta x_S=\Delta y_S=0.3$ a.u., gas density $\mathcal{N}=5.17\times 10^{-5}$ a.u.}\label{Lsq}
\end{figure}

We start by reporting the results of the 1d-2d MS model supplemented by the homogeneous
transport equation \cref{model1_b},  setting first $v_g=c$. The propagation path of the pulse in the gas region $L$, is
divided into $N_1$ subdomains, each containing $N_2$ Maxwell's cells, see
\cref{fig0}. At each Maxwell's time step, the TDSEs are solved several
times (as $\Delta t_M/\Delta t_S\geq10$) to provide the values of ${\bf P}$
in the first cells of each subdomain. The polarization in the remaining cells
of the subdomains is supposed to be computed from the macroscopic transport
equation \cref{model1_b} at the previous time step. The corresponding
parallel computing strategy is presented in \cref{parall_1d_2}. \\
\indent We present in \cref{LP_PE}, the spectral intensity $I(\omega)$ of the
linearly polarized pulse, whose field vector ${\bf E}$ is parallel to the axis of
$\mathrm{H}_2^+$-molecule. The spectral intensity is computed via the Fourier 
transform of the electric field: $I(\omega)=c|\widehat{\bf E}(\omega)|^2/8\pi$, as soon as the pulse envelope escapes from the gas region and appears ``as a whole'' in the last vacuum region.
As we see, when the initial intensity is not high, e.g. $I=5\times10^{13}$ W/cm$^2$, the MS model enriched even by the
simplest evolution equation provides decent approximations of spectra
depending on the chosen partitions of the total number of cells in the gas
region $N$ into number of subdomains $N_1$ and number of cells per each
subdomain $N_2$ (see the legend). \\
\indent Within the same approach we also modeled the propagation of the circularly polarized
initial pulse. In particular, we report in \cref{Lsq} the intensities of
the first generated odd harmonics as functions of the propagation length in
the gas. Comparing this graph with results from \cite{Lyt}, computed within the
``pure'' model, we observe the concordance of the results within $5\%$, even though the new data were
obtained at a much lower cost of computing effort. Indeed, the suggested
approach helps to reduce the computational complexity with respect to the
original model.
\Cref{tab_time} shows that the processing times for the cases of the
``pure'' Maxwell-Schr\"odinger model and the MS model enriched by the
polarization evolution equation are comparable, while in the first case the
number of the engaged processors is up to 16-fold. For example, in the case
of circularly polarized initial pulse, the same spectrum was obtained for the same times but
using 256 processors (``pure'' model) and 16 processors (MS model coupling
with the polarization evolution equation). Note that the computational times
for the circularly polarized-case are higher, as we here used $500\times 500$ points grids to
solve the TDSE, while in LP-case $300\times 300$ points grids are sufficient
to properly computes of the wavefunctions.\\
\begin{table}[h]
\begin{center}
\caption{Processing times for computation of propagation of linearly polarized (LP) and circularly polarized (CP) pulses in the gas, depending on the product of the number of subdomains $(N_1)$ and the number of cells per subdomain $(N_2)$}
\begin{tabular}{|c|c|c|c|c|}
  \hline
   $N_1\times N_2$&64&$8\times 8$&256&$16\times 16$ \\
  \hline\hline
  Proc. time (LP) (h:m)&03:49&03:56&04:35&04:44\\
\hline
  Proc. time (CP) (h:m)&-- &-- &13:35&13:04\\
 \hline
\end{tabular}
\end{center}
\label{tab_time}
\end{table}
\subsection{The MASP Model Enriched by the Nonlinear Evolution Equation}
\begin{figure}[b!]
\centering
\subfloat[]{\label{Ref_dataA}\includegraphics[scale=0.475]{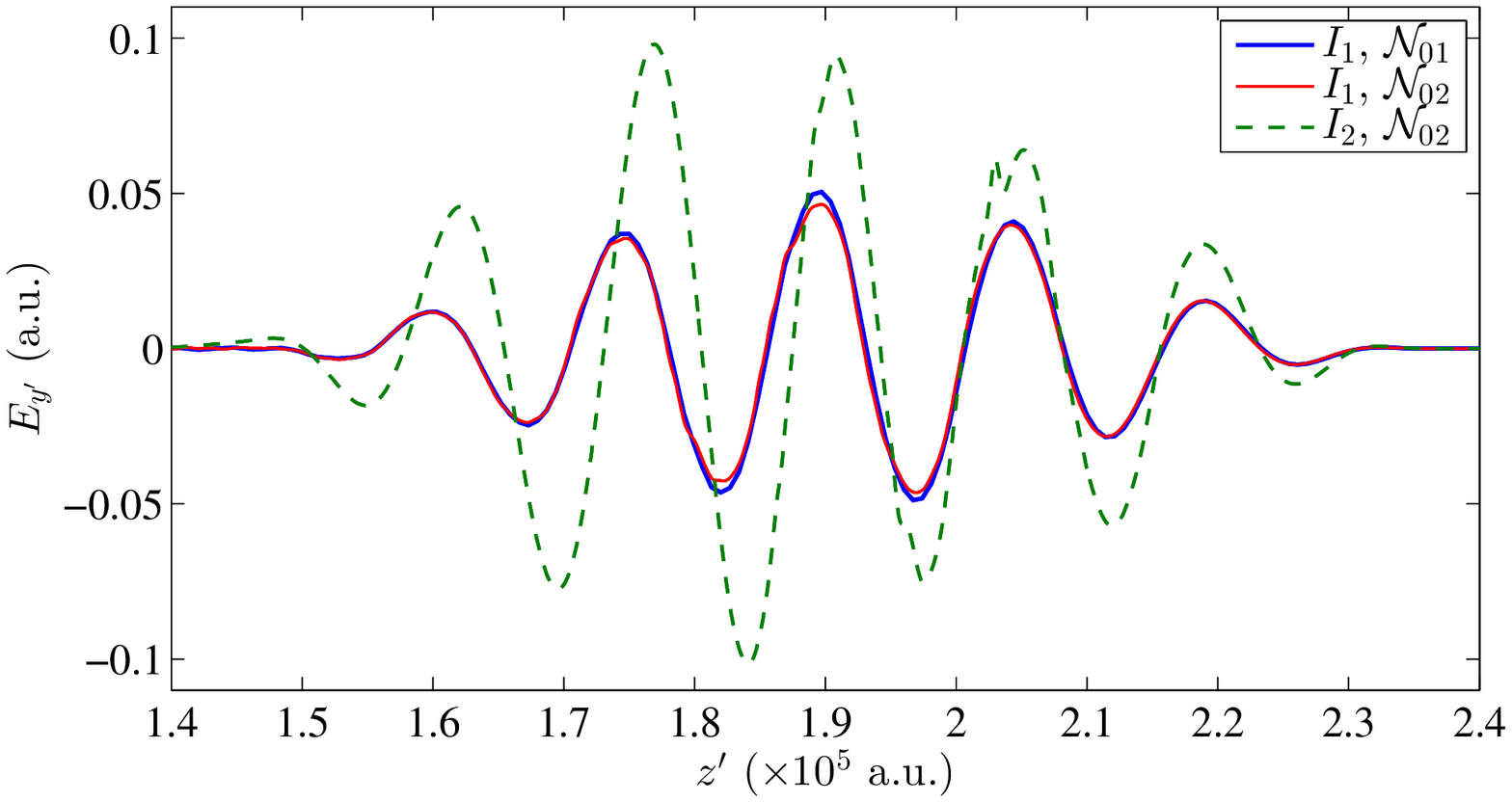}}
\vfill
\subfloat[]{\label{Ref_dataB}\includegraphics[scale=0.475]{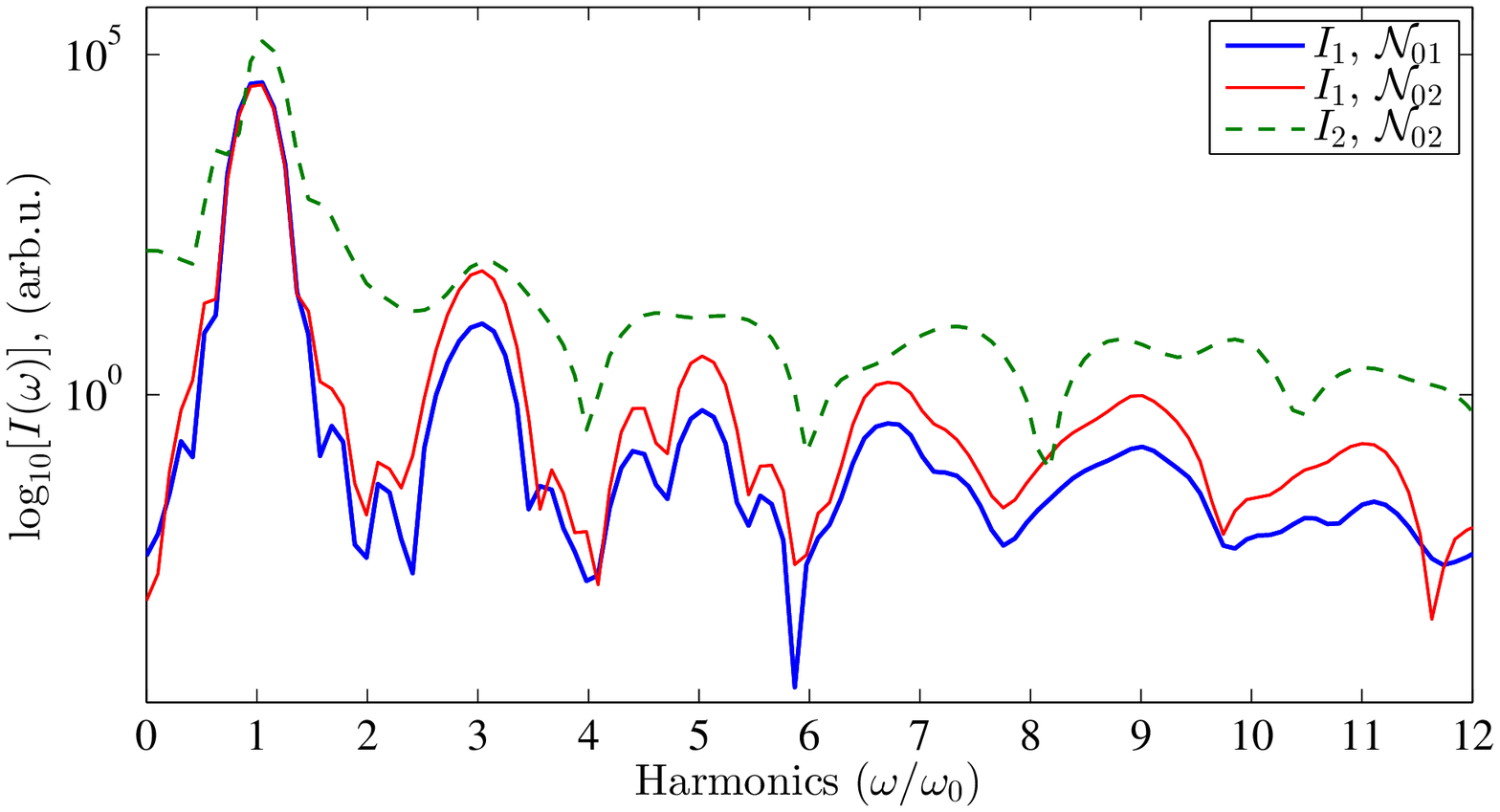}}
\caption{Results of the pure MASP model: a) electric
field as a function of space and b) spectral intensity of the laser
pulse after propagation through the gas region $L=4\times64\times100$ a.u.
Initial pulse intensities: $I_1=10^{14}$ W/cm$^2$, $I_2=5\times10^{14}$ W/cm$^2$,
number density of thee gas: $\mathcal{N}_{01}=1.63\times10^{-5}$ a.u., $\mathcal{N}_{02}=5.17\times10^{-5}$ a.u.}\label{Ref_data}
\end{figure}

\indent In order to simulate the propagation of pulses with higher initial
intensities, we consider the more complicated polarization evolution
equation \cref{N7_1}. Further, we
will compare the transmitted electric fields and spectra computed using the
models enriched by the nonlinear and simple polarization equations with the results of the
``pure'' MS/MASP models, considering the latter as reference data, see
\cref{Ref_data}. Let us briefly discuss these figures. Note that
in our computations we used different intensities of the initial pulse,
$I_1=10^{14}$ W/cm$^2$, $I_2=5\times10^{14}$ W/cm$^2$ and different number
densities $\mathcal{N}_{01}=1.63\times10^{-5}$ a.u. and
$\mathcal{N}_{02}=5.17\times10^{-5}$ a.u. It is clear that the higher
intensity of the pulse and density of the gas, the greater influence of the
free electron currents on dynamics of the process. For example, in case of
initial pulse with $I_1$, $L^2$-norm of the wavefunction is
$\|\psi\|^2_{L^2}\approx0.66$, while in case of $I_2$ we obtained
$\|\psi\|^2_{L^2}\approx10^{-6}$, which results in significant difference in
the free electron number density \cref{Par_ND}. That is why the computations
involving $I_2$ were performed with the MASP model. Recall that unlike the MS
model, the MASP model takes into account the currents of free electrons.

\begin{figure}[p]
\centering
\subfloat[]{\label{Corrections_I1A}\includegraphics[scale=0.7]{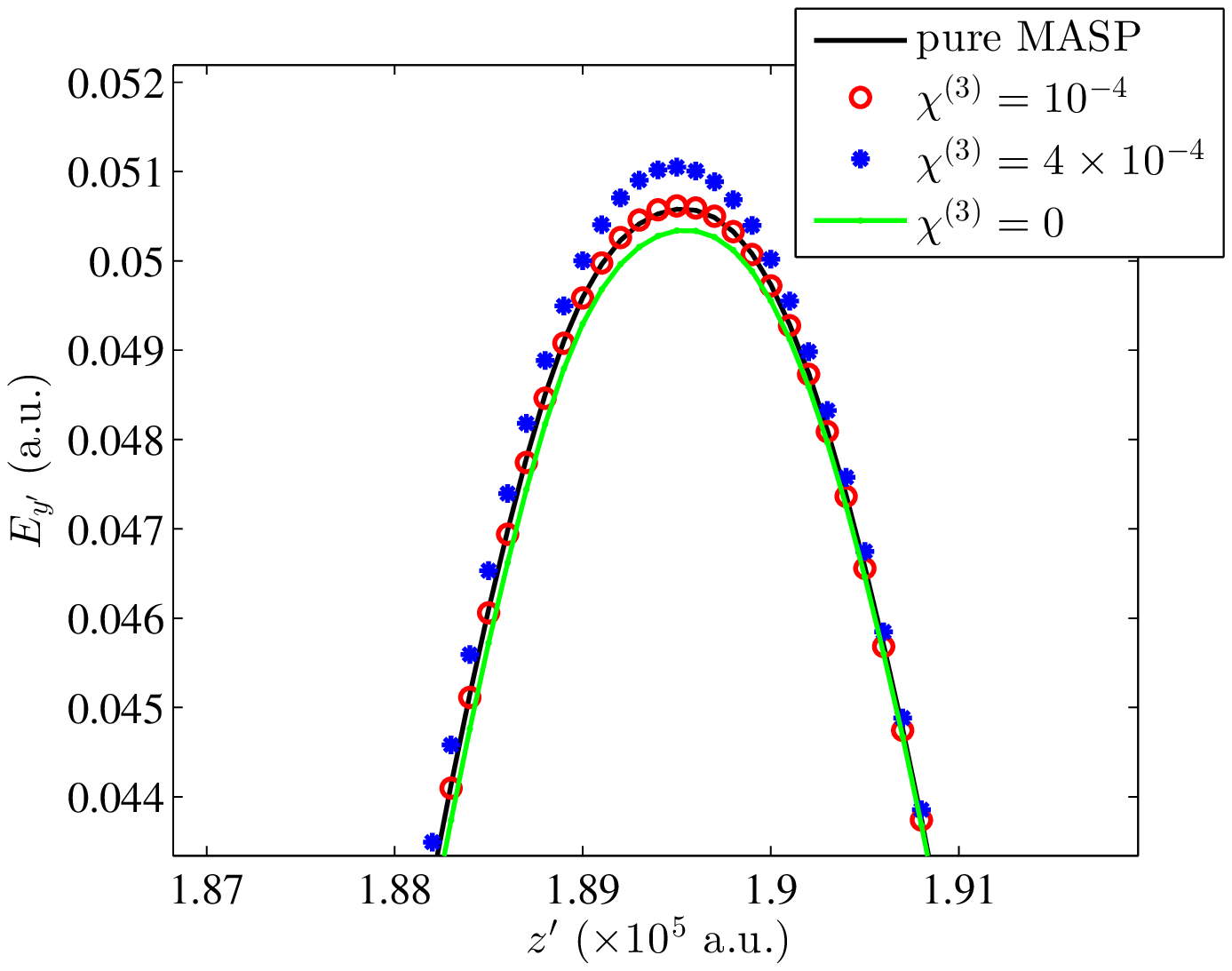}}
\vfill
\subfloat[]{\label{Corrections_I1B}\includegraphics[scale=0.7]{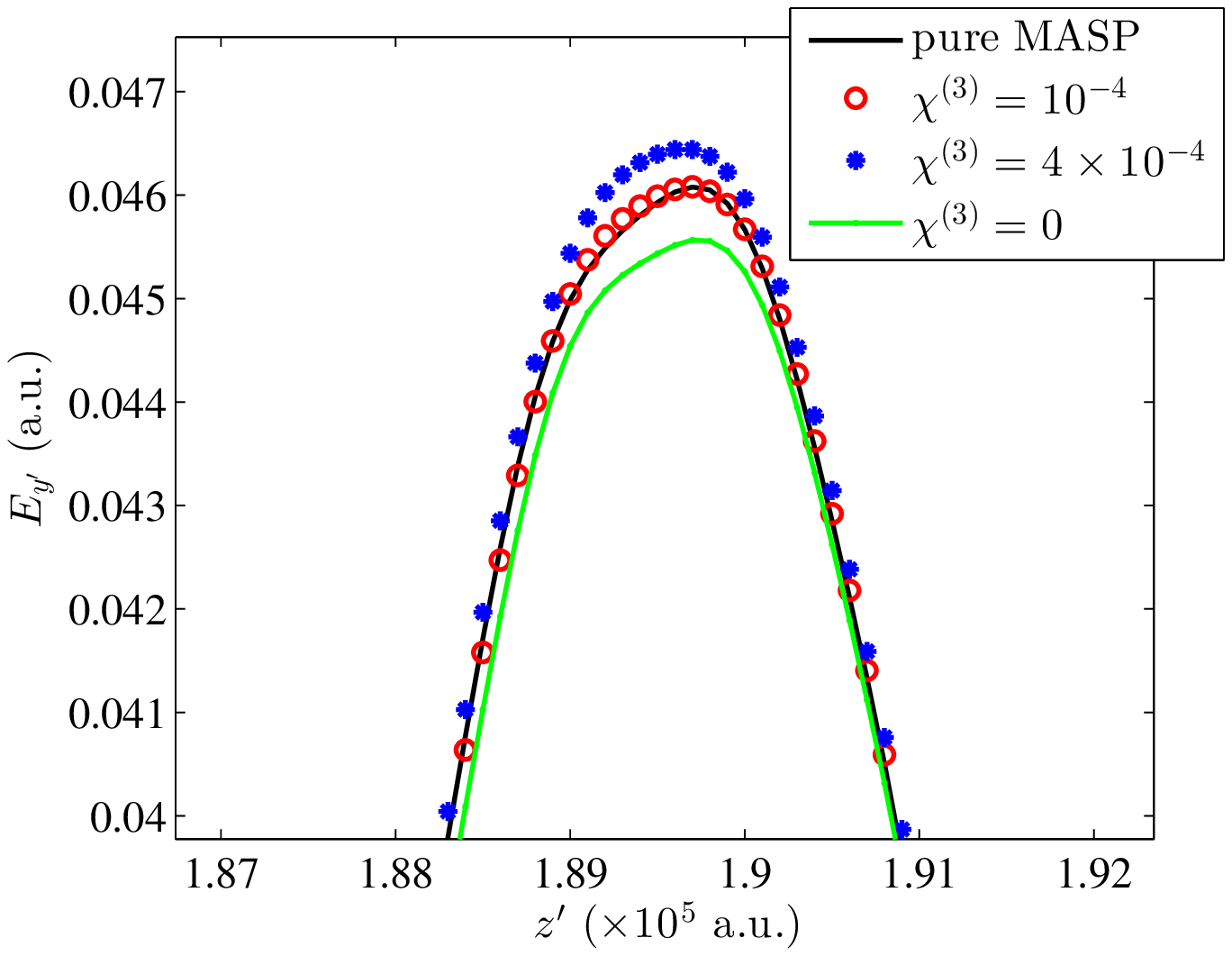}}
\caption{Results of the MASP model supplemented by the
polarization evolution equation in comparison with the results of the ``pure'' MASP model:
transmitted electric fields for initial intensity $I_1=10^{14}$ W/cm$^2$ and
gas number density a) $\mathcal{N}_{01}=1.63\times10^{-5}$ a.u.,
b) $\mathcal{N}_{02}=5.17\times10^{-5}$ a.u. Propagation length in gas $L=256\times100$ a.u.,
which in case of engaging the evolution equation is
divided into 4 subdomains each containing 64 cells. Linear instantaneous
susceptibility for polarization equation $\chi^{(1)}=1.83\times10^{-5}$,
while the coefficient $\chi^{(3)}$ is a model parameter, see the legends.}
\label{Corrections_I1}
\end{figure}

\begin{figure}[h!]
\center{\includegraphics[scale=0.7, trim=45 0 20 0, clip]{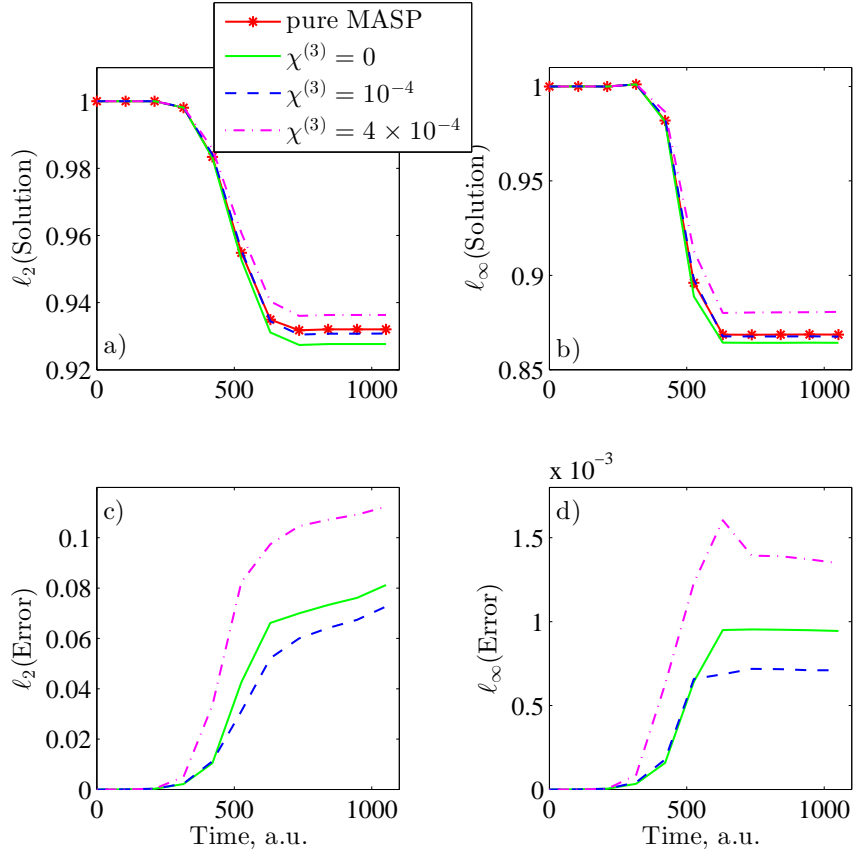}}
\caption{a) $\ell_2$-norms of solutions presented in \cref{Corrections_I1B}; b) $\ell_\infty$-norms of the same solutions;
c) $\ell_2$-norms of errors of these solutions with respect to the MASP model results;
d) $\ell_\infty$-norms of errors of these solutions with respect to the MASP model results.}
\label{LnorI1D2}
\end{figure}

\begin{figure}[p]
\centering
\subfloat[]{\label{Spectra_I1A}\includegraphics[scale=0.6]{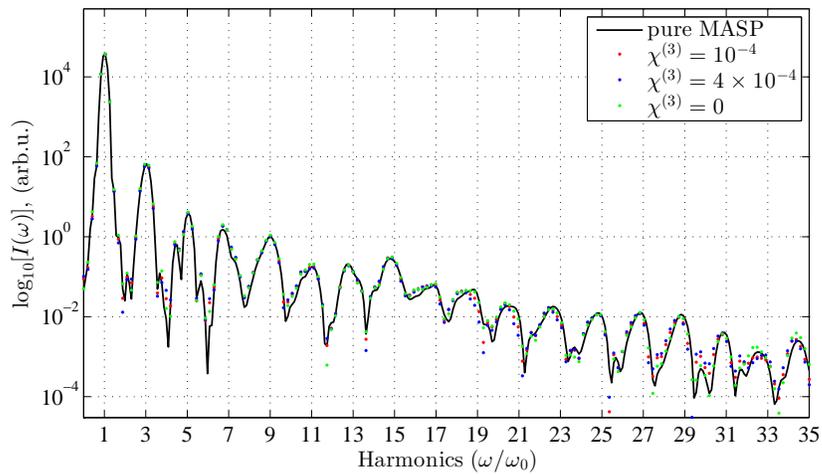}}
\vfill
\subfloat[]{\label{Spectra_I1B}\includegraphics[scale=0.6]{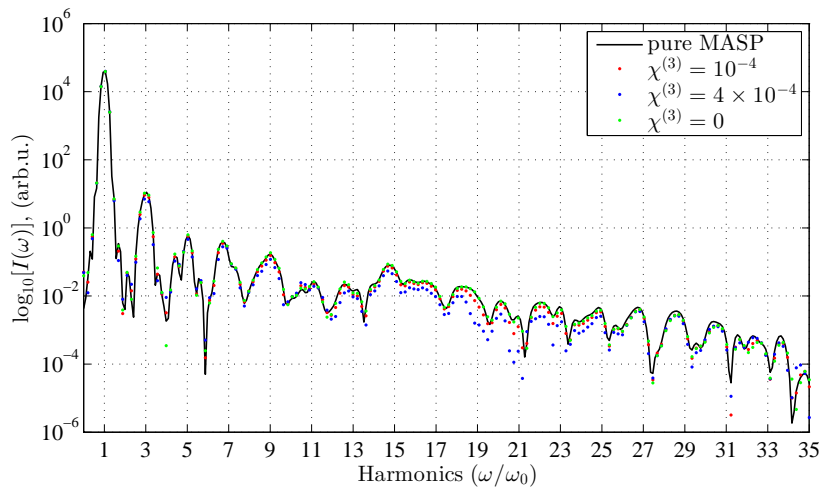}}
\caption{Spectral intensities of high harmonics. The same parameters as for \cref{Corrections_I1A,Corrections_I1B}.}\label{Spectra_I1}
\end{figure}

\begin{figure}[p]
\centering
\subfloat[]{\label{Field_I2A}\includegraphics[scale=0.6]{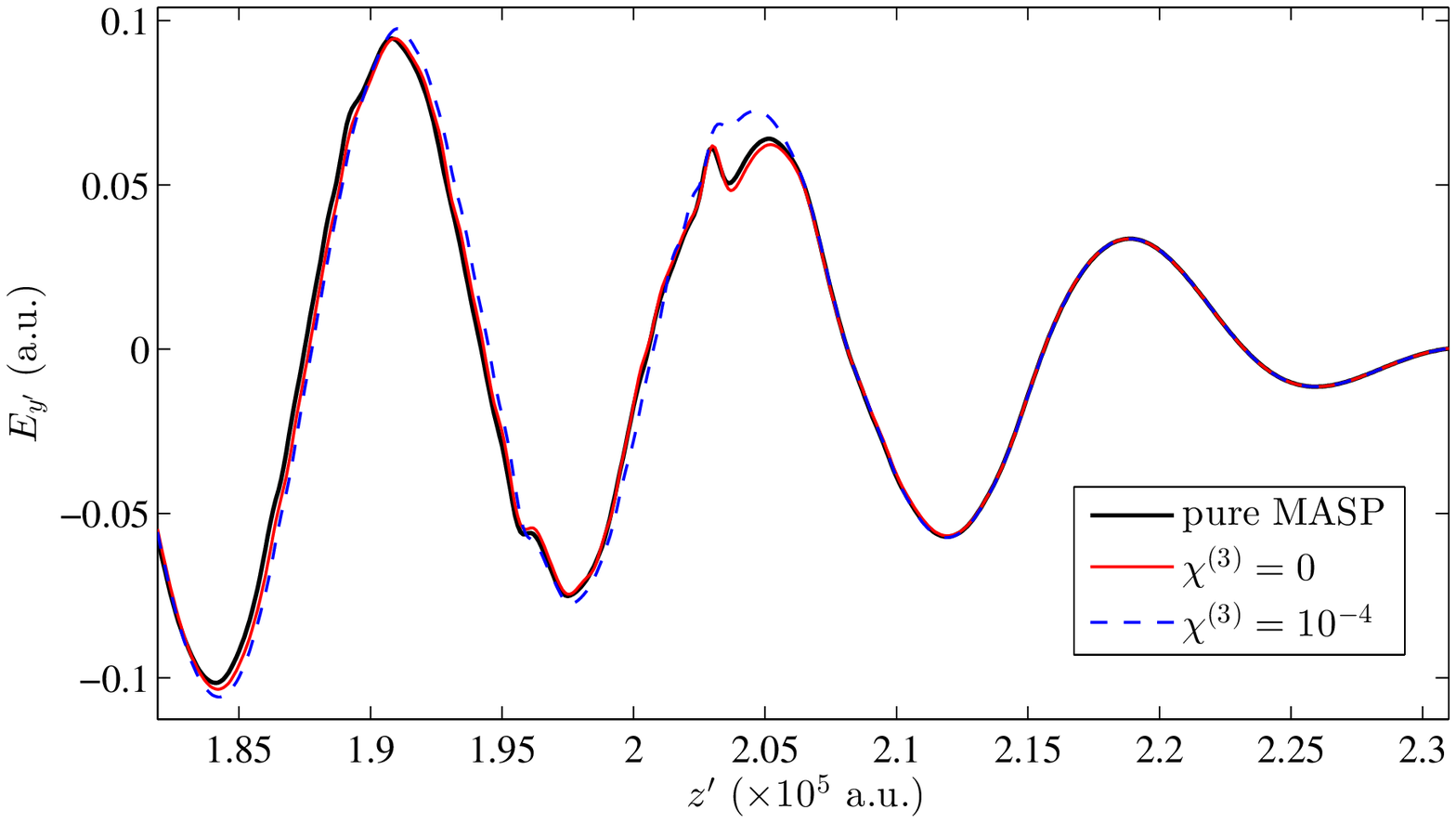}}
\vfill
\subfloat[]{\label{Spectra_I2B}\includegraphics[scale=0.6]{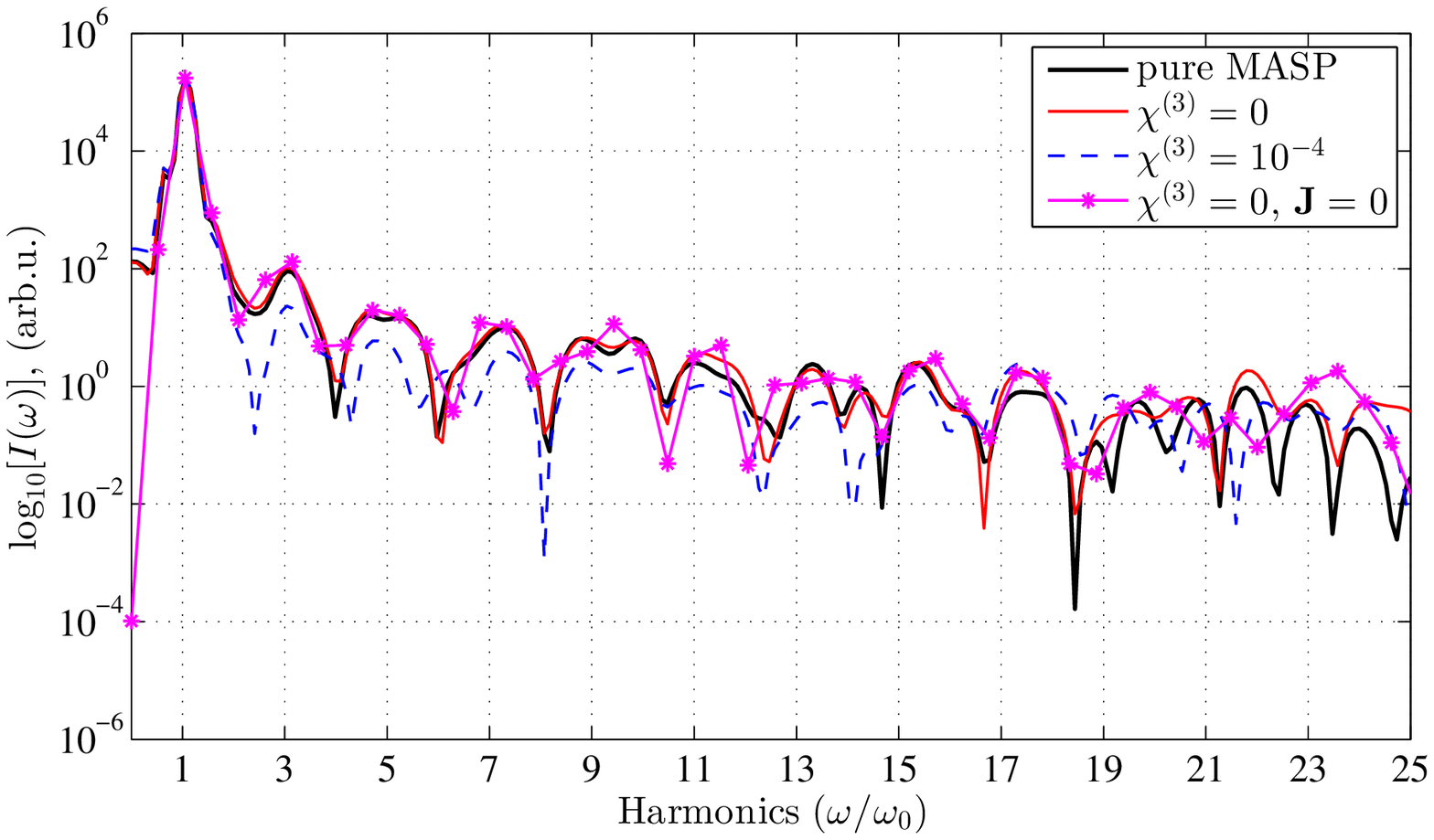}}
\caption{Results of the MASP model supplemented by the polarization evolution equation in comparison
with the results of the ``pure'' MASP model: (a)
transmitted electric fields for initial intensity $I_2=5\times10^{14}$ W/cm$^2$
and gas number density $\mathcal{N}_{02}=5.17\times10^{-5}$ a.u.
Path in a gas $L=256\times100$ a.u., divided into 4 subdomains each containing
64 cells, $\chi^{(1)}=1.83\times10^{-5}$,
$\chi^{(3)}$ is a model parameter, see the legends; (b) Spectral Intensity for solutions shown in panel (a), additional solution computed in case ${\bf J}=0$.}\label{Corrections_I2}
\end{figure}

\begin{figure}[t!]
\centering
\subfloat[]{\label{LongPropSpecA}\includegraphics[scale=0.59]{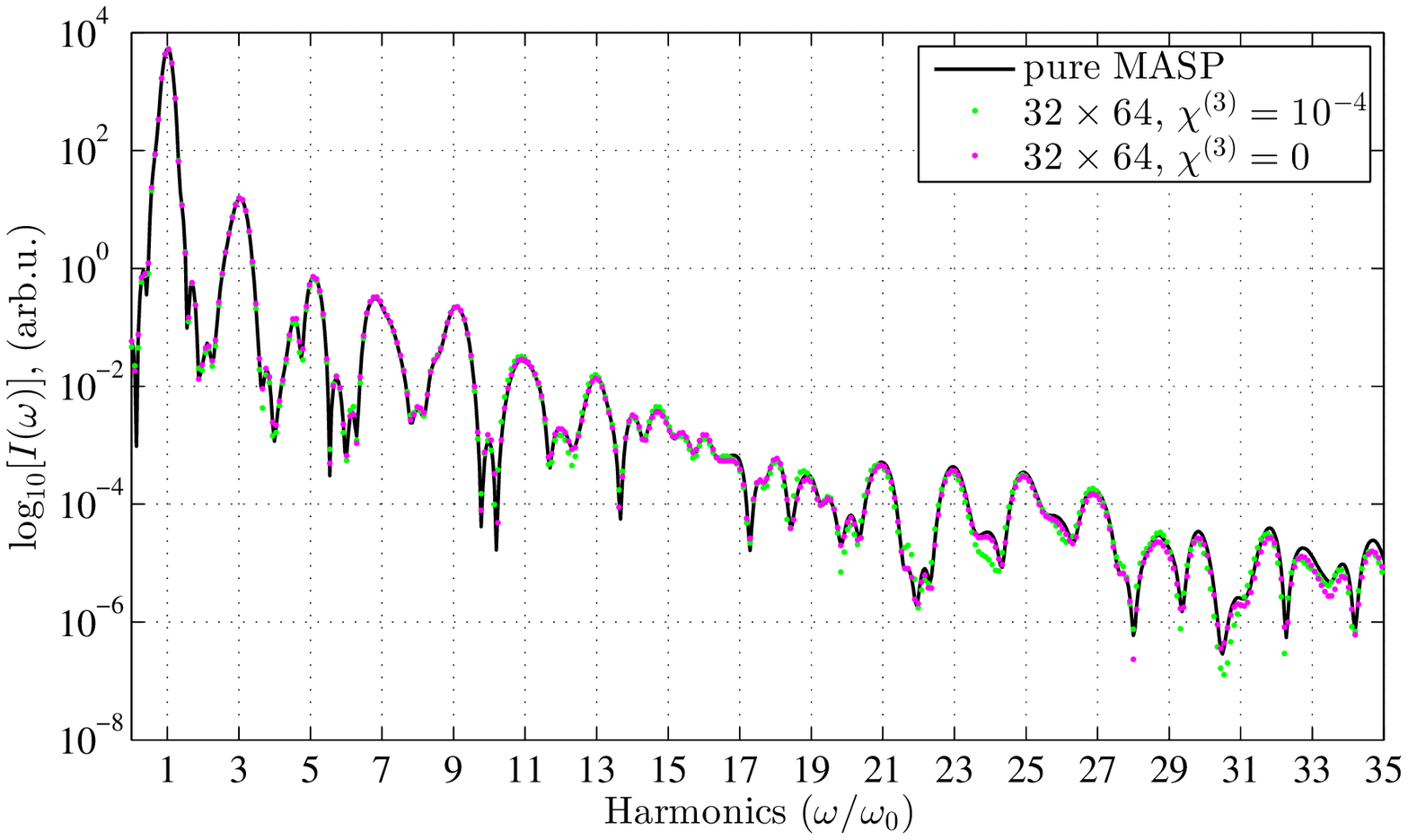}}
\vfill
\subfloat[]{\label{LongPropSpecB}\includegraphics[scale=0.59]{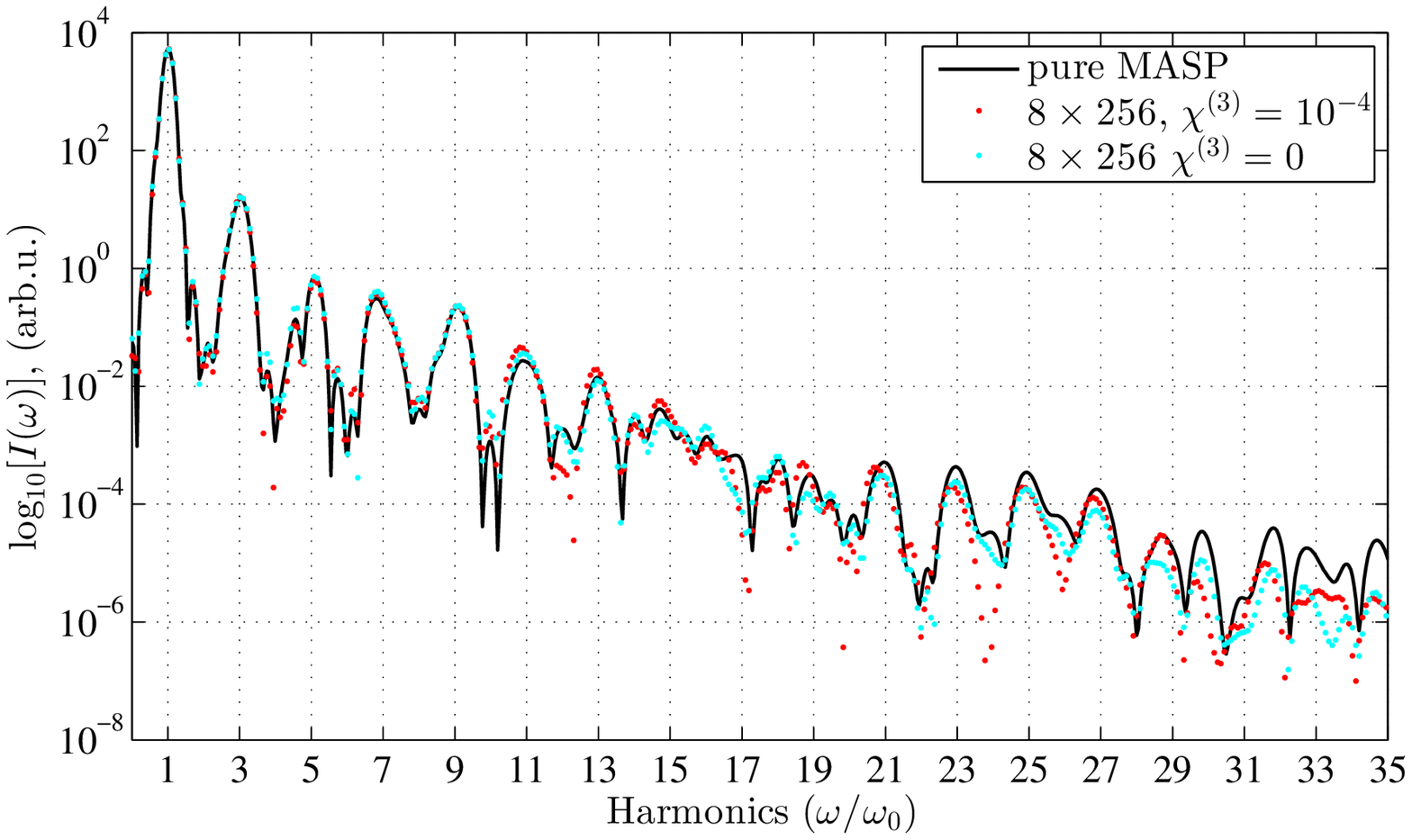}}
\caption[Spectra of high harmonics: different domain decompositions.]
{Spectra of high harmonics obtained within the ``pure'' MASP model
and using the evolution equation for polarization. The pulse initial
intensity $I_1=10^{14}$ W/cm$^2$, gas density $\mathcal{N}_{01}=1.63\times10^{-5}$
a.u., propagation length in a gas $L=204800$ a.u. (0.01 mm), which is
divided into a) 32 subdomains each containing 64 cells and b) 8 sundomains each
containing 256 cells.} \label{LongPropSpec}
\end{figure}

\indent As we see from \cref{Ref_dataA,Ref_dataB}, if we consider pulses with the
same initial intensity $I_1$, but propagating in gases of different
density, for definiteness $\mathcal{N}_{01}<\mathcal{N}_{02}$, the amplitude
of the transmitted electric field is a bit lower in the case of a higher
density,  due to stronger ionization losses. However, the generation
of the high odd harmonics is more intense for $\mathcal{N}_{02}$, see \cref{Ref_dataB}, as the polarization of the medium is proportional to the gas number
density. On the other hand, in the case of initial intensity $I_2>I_1$ the
level of ionization becomes much higher (with respect to initial intensity
$I_1$). As a result, the amplitudes of the high harmonics are relatively smaller. \\
\indent Including the instantaneous susceptibility coefficient $\chi^{(3)}$
(or even $\chi^{(5)}$, $\chi^{(7)}$ and so on) as a parameter in the
polarization evolution equation \cref{N7_1}, we are pursuing two objectives:
to describe more accurately (i) the electromagnetic field profile, and (ii) the spectrum of
harmonics. In the following set of figures, we report the data computed within
the MASP model supplemented by the evolution equations. We start from the
length of the gas region $L=256\times100$ a.u., with propagation time $T=1051.87$ a.u. We use 3 different values of
$\chi^{(3)}$ to test the model. At first glance, even simulations engaging
the transport equation with $v_g=c$ and $\chi^{(3)}=0$, describe the field
profile decently; however, as we are comparing the corrections for nonlinear
effects accumulating very slowly, we need to pay attention to the very top
parts of the electric field profiles, see \cref{Corrections_I1}. Then we
notice with initial intensity~$I_1$, the parameter $\chi^{(3)}=10^{-4}$ works better for both gas densities. In
\cref{LnorI1D2}, we also report $\ell_2$ and
$\ell_\infty$-norms of the solutions presented in
\cref{Corrections_I1B} and the errors computed with respect to the
``pure'' MASP model. As we observe, in all the figures the dashed blue curves corresponding to
$\chi^{(3)}=10^{-4}$ provides the closest agreement with the results of
the ``pure'' MASP model in conformity with results in \cref{Corrections_I1}. \\
\indent Now we proceed to the presentation of the high harmonics spectra, see
\cref{Spectra_I1}. The simulation parameters for these spectra are
exactly the same as for \cref{Corrections_I1}. We observe that the selection
$\chi^{(3)}=4\times10^{-4}$ a.u. yields less precise results than the two other options.
In this case, we are overestimating again the response of the bound electrons.
Moreover, the evolution equation \cref{N7_1} with $\chi^{(3)}$ in the RHS, may account corrections to polarization of the 3$^{rd}$ harmonic only, while these computations seems to be adequate even with homogeneous transport equation \cref{model1_b}.\\
\indent The situation changes at higher initial intensities, e.g.
$I_2=5\times10^{14}$ W/cm$^2$ that, as we know, results in significant
ionization. In this case, introducing instantaneous nonlinear susceptibility
does not seem appropriate. In \cref{Field_I2A}, we observe the differences between the solutions of the ``pure'' and ``enriched'' MASP models. This was expected, as $\chi^{(3)}$ describes the response of bound electrons, whose number is way below the number of free electrons at high intensity. Even
so, we still can rely on simulation with the transport equation as it follows
from comparison between the heavy black and the thin red curves in
\cref{Field_I2A}.\\
\indent In \cref{Spectra_I2B} we report the high harmonic spectra formed
after propagation of the pulse of initial intensity $I_2=5\times10^{14}$
W/cm$^2$ through the gas. The first three solutions (according to the legend)
correspond to the electric field profiles represented in \cref{Corrections_I2}. The 4$^\mathrm{th}$ solution was obtained under the assumption ${\bf J}=0$, which as we know is wrong for that intensity.
In the large, one can observe the best convergence, inter alia,  between solutions of the pure MASP model (heavy black curves on \cref{Field_I2A,Spectra_I2B}) and those of the MASP model enriched by the homogeneous transport equation (thin red curves). Just as expected (see the dashed blue curve), adding the perturbative term with $\chi^{(3)}$ to the polarization
equation is unsuitable in such ionized gas. We also deduce that including the
free electron currents in the model is essential, as the solution in magenta, see 
\cref{Spectra_I2B}, is not a good candidate to fit the ``pure'' MASP model solution. \\
\indent Regarding the accurate simulations of the high harmonics spectra within the enriched model, the most important factor (than nonlinearity in the polarization equation) is how Maxwell's domain is decomposed in subdomains, see \cref{LongPropSpecA,LongPropSpecB} where the propagation length is
$L=204800$ a.u. Say, decomposition into 32 subdomains each containing 64 cells, allows us to
simulate accurately up to 35 harmonics, while spending less computing
resources: 94~h$\times$256~proc vs. 12~h$\times$32~proc. Moreover, if we are
interested in the first 11 harmonics only, we can use
decomposition 8$\times$256, which takes 12~h$\times$8~proc.
\clearpage

%% file: conclusion.tex
\section{Conclusion}\label{sec:conclusion}
We have derived and studied some extensions of the MASP model \cite{cicp,phyd}, allowing for a wider range of application of this generic nonperturbative model. We have demonstrated that including a polarization evolution equation to the MASP model allows for a significant reduction of the overall computational cost of the original model. We found out that the most universal type of this equation is the homogeneous transport equation. More complex nonlinear polarization equations can be useful when the level of ionization is moderate. By contrast, at high intensity (resulting in high level of ionization), the cheapest choice is the homogeneous transport equation, especially for simulating high order harmonic spectra. In this case, it is shown that the current of free electrons must also be included in the model.